\documentclass[a4paper,12pt]{amsart}
\usepackage[left=1in, right=1in, top = 1.5in, bottom = 1.5in]{geometry}
\usepackage[round]{natbib}
\usepackage[utf8]{inputenc}
\usepackage[all]{xy}
\usepackage{amssymb}
\usepackage{amsmath}
\usepackage{csquotes}
\usepackage{subfig}
\usepackage{enumerate}
\usepackage{graphicx}
\usepackage{enumerate}
\usepackage{amsaddr}
\usepackage{color}
\usepackage{mwe}
\usepackage{appendix}
\usepackage{centernot}
\usepackage{mathtools}
\usepackage{comment}
\usepackage[ruled,vlined]{algorithm2e}
\DeclareMathOperator*{\eff}{eff}

\usepackage{soul}
\usepackage{xcolor}
\def\@centernot#1#2{%
  \mathrel{%
    \rlap{%
      \settowidth\dimen@{$\m@th#1{#2}$}%
      \kern.5\dimen@
      \settowidth\dimen@{$\m@th#1=$}%
      \kern-.5\dimen@
      $\m@th#1\not$%
    }%
    {#2}%
  }%
}
\makeatother

\newcommand{\independent}{\perp\mkern-9.5mu\perp}

\usepackage{tikz}
\usetikzlibrary{arrows.meta,shapes}
\usetikzlibrary{arrows,shapes.arrows,shapes.geometric,shapes.multipart,
decorations.pathmorphing,positioning,shapes.swigs,}

\usepackage{setspace}
\newtheorem{lemma}{Lemma}
\newtheorem{proposition}{Proposition}
\newtheorem{corollary}{Corollary}
\theoremstyle{definition}
\newtheorem{assumption}{Assumption}
\newtheorem{definition}{Definition}
\newtheorem{example}{Example} %[section]
\theoremstyle{remark}
\newtheorem{remark}{Remark}
\newtheorem*{proposition*}{Proposition}

\MakeOuterQuote{"}\EnableQuotes
\DeclareUnicodeCharacter{00A0}{ }

\title[]{Optimal regimes for algorithm-assisted human decision-making}
\author{Mats J. Stensrud \& Julien Laurendeau \& Aaron L. Sarvet} \address{
Department of Mathematics, Ecole Polytechnique Fédérale de Lausanne, Switzerland 
} %$^1$  
\begin{document}
\maketitle
%individualized treatment 
\begin{abstract}
We consider optimal regimes for algorithm-assisted human decision-making. Such regimes are decision functions of measured pre-treatment variables and, by leveraging natural treatment values, enjoy a ``superoptimality'' property whereby they are guaranteed to outperform conventional optimal regimes. When there is unmeasured confounding, the benefit of using superoptimal regimes can be considerable. When there is no unmeasured confounding, superoptimal regimes are identical to conventional optimal regimes. Furthermore, identification of the expected outcome under superoptimal regimes in non-experimental studies requires the same assumptions as identification of value functions under conventional optimal regimes when the treatment is binary. To illustrate the utility of superoptimal regimes, we derive new identification and estimation results in a common instrumental variable setting. We use these derivations to analyze examples from the optimal regimes literature, including a case study of the effect of prompt intensive care treatment on survival. 
\end{abstract}

\section{Introduction}

Foundational work on causal inference and dynamic treatment regimes presents a promising pathway towards precision medicine \citep{robins1986new,richardson2013single,murphy2003optimal, robins2004optimal,tsiatis2019dynamic,kosorok2021introduction}. In a precision-medicine system, decision rules might be algorithmically individualized based on an optimal rule previously learned from non-experimental or experimental data \citep{topol2019high}. However, wide-scale implementation of such a system will usually roll-out under the supervision of existing medical care providers \citep{matheny2019artificial}. Indeed, there is some resistance to the notion that implementation of an optimal regime, successfully learned from the data, will result in better expected outcomes on average, compared to existing human-decision rules. This resistance stems in part from the belief that existing care providers often will have access to relevant information for decision-making that is not recorded in the observed data \citep{verghese2018computer}. While these beliefs do not prohibit identification of decision rules that are optimal with respect to a set of measured covariates \citep{cui2021necessary,cui2021semiparametric,qiu2021optimal,pu2021estimating,han2021comment,miao2018identifying, qi2021proximal,kallus2021minimax}, care providers may be inclined to override the treatment recommendations provided by the identified optimal regimes, based on their privileged patient observations. 

We present methodology for leveraging human intuition, given by the intended treatment values, by identifying a \textit{super}optimal regime using data from either non-experimental or experimental studies, and clarify when a fusion of such data is beneficial. The superoptimal regime will indicate to a care provider --- in an algorithm-assisted decision setting --- precisely when expected outcomes would be maximized if the care provider overrides the optimal regime recommendation and, importantly, when the optimal regime recommendation should be followed regardless of the care-provider's assessment. This superoptimal regime is identical to the (conventional) optimal regime in settings with no unmeasured confounding. However, when there is unmeasured confounding, the superoptimal regime yields as good or better expected outcomes compared to both the optimal regime \textit{and} the implicit (factual) regime independently implemented by the care provider in the observed data, which have been studied in previous work \citep{cui2021necessary,cui2021semiparametric,han2021comment,miao2018identifying, qi2021proximal,kallus2021minimax}. Furthermore, in many settings identification of the superoptimal regime requires \textit{no additional assumptions} beyond those used to identify the optimal regime and its expected outcome, i.e.\ value function, allowing us to identify superoptimal regimes by making small modifications of existing methods \citep{ cui2021necessary,cui2021semiparametric,qiu2021optimal}. 

This work builds on literatures arising from historical interest in the the so-called average treatment on the treated (ATT) \citep{bloom1984accounting, heckman1990varieties}. One strand of literature expands on the ATT by defining, identifying and estimating a general class of causal parameters defined by the values of patients' natural treatment choices or intentions in the absence of intervention \citep{robins2004effects, robins2006comment, haneuse2013estimation, richardson2013single, young2014identification, diaz2021nonparametric}. While this strand of literature is ostensibly interested in the values of ATT-like parameters \textit{per se}, a second strand of literature is especially concerned with heterogeneity among them and the implications for transportability of clinical trial results. This second strand focuses on identification of ATT-like parameters using augmented experimental designs, sometimes referred to as patient preference trials. Unlike conventional two-arm randomized trials, patient preference trials include an additional (third) arm where individuals can choose the treatment they receive \citep{knox2019design}. While recruitment to the third arm historically has been done in different ways \citep{collinge2009safety,mclaughlin2022algorithmic,rucker1989two}, modern formulations of patient preference trials require that individuals state their treatment preference before randomization to either treatment, control, or taking their (stated) preferred treatment \citep{long2008causal, knox2019design}. The third arm corresponds to an observational setting in the sense that a representative sample of individuals selects treatment based on their own preferences. If these preferences reflect knowledge or belief about the response to treatment, then knowing the preferences gives information that is relevant to making decisions. However, this information is inaccessible in conventional randomized trials. We can view patient preference trials as target trials \citep{hernan2016using} that would allow the identification of our parameters of interest by design.

Finally, an independent collection of contributions in the machine learning literature studies the optimal selection of treatment based on a patient's treatment intentions in an online experimental learning setting \citep{bareinboim2015bandits, forney2017counterfactual, forney2019counterfactual}. 

%These related literatures represent an evolution in consideration of parameters conditional on natural treatment choices, but have largely developed independently. 

Our contribution unifies and extends these related, but independently developed, literatures and our framing clarifies connections between optimal and superoptimal regimes that are obscured in the extant literature. Furthermore, like the literature characterized by \citet{robins2004effects}, \citet{richardson2013single} and others, we do not focus on, or restrict ourselves to, experimental settings: we emphasize the non-experimental setting and locate the results on experimental settings as special cases.

%Previous works from different domains have independently suggested augmented experimental designs , called 'counterfactual randomization' \citep{forney2019counterfactual} or 'patient preference trials' \citep{knox2019design}, where the \textit{intended} treatment decisions are measured and used to identify effects conditional on the intended treatment. Along the same lines,

%A related literature focuses on treatment effect heterogeneity based on \textit{ex post facto} treatment choice (e.g. between the average treatment effect among those actually treated and untreated) or \textit{a priori} treatment intentions  

%The remainder of this article is organized as follows. In Section \ref{sec: prelim}, we describe the observed data structure, introduce potential outcome variables and define superoptimal regimes. In Section \ref{sec: superoptinal regimes}, we derive optimality guarantees and identification results on superoptimal regimes. In Section \ref{sec: case study iv}, we evaluate outcomes under superoptimal regimes examples, inspired by previous work in the optimal regimes literature. In Section \ref{sec: icu example} we analyze the ICU data.  In Section \ref{sec: discussion}, we suggest directions for future research. 

\section{Preliminaries}
\label{sec: prelim}
\subsection{Non-experimental data structure}
Consider a treatment $A \in \{ 0,1 \}$, a pretreatment vector $L \in \mathcal{L} $, and an outcome $Y \in \mathbb{R}$. 
Suppose that we have access to $n$ iid observations of $(L,A,Y)$ among patients who received treatment in a non-experimental setting. An unmeasured variable $U \in \mathcal{U}$ can be a common cause of $A$ and $Y$. Some of our results, in particular those in our case study in Section \ref{sec: case study iv}, will further rely on observations of an instrumental variable (IV) $Z \in \{0,1\}$, and we use $\mathcal{O}=(Z,L,A,Y)$ to denote the observed non-experimental data in the IV setting.

\subsection{Potential outcomes and the natural values of treatment}
\label{sec: pot out and nat tr}
Let superscripts denote potential outcome variables. In particular, $Y^a$ is the potential outcome when treatment $A$ is fixed to the value $a\in\{0.1\}$. More specifically will let $Y^g\equiv Y^{g(V)}$ be the potential outcome under an arbitrary regime $g$, where the treatment is assigned as a function of measured covariates $V \subset \mathcal{O}$. Following \citet{richardson2013single}, we use the $+$ symbol to distinguish between the assigned value of treatment under the regime ($A^{g+}$) and the \textit{natural value} of the treatment under the regime ($A^{g}$). The natural value will be important in our arguments, and we state its definition explicitly \citep{richardson2013single}.\footnote{See also \citet{robins2006comment,geneletti2011defining,young2014identification} for interesting discussions on dynamic treatment regimes that depend on the natural value of treatment.}

%$Y^g$ is the potential outcome under an arbitrary regime $g$, where the treatment $a$ is assigned based on measured covariates.

\begin{definition}[Natural value of treatment] \label{def: natval}
The natural value of treatment $A^g$ is the value of treatment that an individual would choose in the absence of it being assigned by an intervention.
\end{definition}

We have used counterfactuals to define the natural values of treatment, like previous authors \citep{haneuse2013estimation,young2014identification,munoz2012population}. However, we could alternatively give the natural values an interventionist interpretation, which does not require conceptualization of counterfactuals: following \citet{robins2006comment} and \citet{geneletti2011defining}, the natural value of treatment is a variable that is temporally prior but deterministically equal to the active treatment in non-experimental data; that is, the natural treatment value and the active treatment value are equal with probability one. 

The natural value of treatment under the regime $g$, $A^g$, is equal to $A$ in any non-experimental study that investigates the effect of a point treatment, that is, $A = A^g$ w.p.1. Thus, if $A$ is observed, then $A^g$ is observed. In particular, this is true in non-experimental studies that identify causal effects in the presence of unmeasured confounding, e.g.\ using IVs or proxy variables \citep{miao2018identifying,tchetgen2020introduction}. %Henceforth, we will simply denote the natural value of treatment with $A$ because we focus on a point treatment setting.

To fix ideas about natural treatment values, consider a doctor who determines whether a patient will be transferred to an intensive care unit (ICU); let $A=1$ denote ICU admission and $A=0$ denote no ICU admission. In the observed data, the doctor \textit{determined} the ICU admission and thus the natural value $A^g$ is equal to $A$ with probability 1. We could, however, conceive a regime where the assigned ICU admission, $A^{g+}$, is determined by some arbitrary function $g$ of patient characteristics such as the pretreatment covariates $L$. It is possible that the assignment $A^{g+}$ differs from the natural value $A$.

\subsection{Definitions of treatment regimes}
\label{sec: def regimes}
In this section we formally define $L$-optimal and $L$-superoptimal regimes in a point treatment setting, where we employ the prefix "$L-$" to emphasize their definitional dependence on the elements in the covariate vector $L$. Throughout, we suppose that larger values of $Y$ are desirable.

\begin{definition}[$L$-optimal regimes]
\label{def: opt}

The $L$-optimal regime, $g_{\mathbf{opt}}$, assigns treatment $A^{g_{\mathbf{opt}}+}=a$ given a vector $L=l$ by
$$
   g_{\mathbf{opt}}(l)  \equiv \underset{a \in \{0,1\} }{\arg \max } \  \mathbb{E}(Y^a \mid L = l) \label{eq: optimal g given l}. %, \text{ where $\mathcal{G}$ is the set of regimes that satisfy impartiality.}   %\nonumber \\
   $$
\end{definition}

\begin{definition}[$L$-superoptimal regimes]
\label{def: supopt}
The $L$-superoptimal regime, $g_{\mathbf{sup}}$ assigns treatment $A^{g_{\mathbf{sup}}+} = a$ given $A=a'$ and $L=l$ by 
 $$
  g_{\mathbf{sup}}(a',l)  \equiv \underset{a \in \{0,1\} }{\arg \max } \  \mathbb{E}(Y^{a} \mid A=a', L = l) \label{eq: optimal g given a l}. %, \text{ where $\mathcal{G}$ is the set of regimes that satisfy impartiality.}   %\nonumber \\
$$
\end{definition}
We denote the counterfactual expectation $\mathbb{E}(Y^{a} \mid L = l)$ as a 'conditional value function'. In particular  $ \mathbb{E}(Y^{g_{\mathbf{opt}}} \mid L = l) $ and  $ \mathbb{E}(Y^{g_{\mathbf{sup}}} \mid A=a', L = l)$ are conditional value functions under the $L$-optimal and $L$-superoptimal regimes, respectively. 

Treatment rules given by $L$-optimal and $L$-superoptimal regimes can be presented in Single World Interventions Graphs (SWIGs) \citep{richardson2013single}, as illustrated by the IV setting in Figure \ref{fig:superoptimal}: the green arrow encodes regime-specific effects of the measured covariates $L$ on the assigned value of treatment under the regime, $A^{g+}$, a feature of both the $L$-optimal and the $L$-superoptimal regime. The blue arrow further encodes the effect of the natural value of treatment $A$ on $A^{g+}$, a feature of the $L$-superoptimal, but not the $L$-optimal, regime.

%, as studied in \citep{qiu2021optimal,cui2021semiparametric}

Consider again the setting where a patient might be transferred to an ICU. Suppose we have access to non-experimental data from a setting where physicians determined ICU admission, and thus $A = A^g$. Using these data, an investigator aims to find the dynamic regime for ICU admission that gives the highest 7 day survival in a future decision setting. To specify this regime, we could assign $A^{g+}$ as a function of measured covariates $L$, describing the patient's age, gender and a collection of clinical measurements. However, beyond using the values of $L$, we could also ask the treating physician the following question: "if you were to choose, would you transfer the patient to an ICU?" The answer to this question would encode the natural treatment value $A$, and we can indeed use both $L$ and $A$ as input to our decision rule; a superoptimal regime will precisely let $A^{g+}$ be a function of both $L$ and $A$.\footnote{Under the assumption that the doctor's response to the question actually agrees with the decision they had made if we did not intervene.} 

This brief example suggests how the natural treatment value interventions feasibly can be implemented; just before intervening we ask the decision maker about the treatment they \textit{intend} to provide, and then we record their response to this question as a covariate. Nearly identical measurement strategies for ``patient preference'' or ''intent'' are leveraged in the literature on patient preference trials (see, for example, \citet{rucker1989two, long2008causal, knox2019design}).
%$\footnote{Unlike certain natural value interventions that have been suggested in the literature, our rules can be implement.} 

We consider identification of superoptimal regimes from observational data, wherein unmeasured confounding between the treatment and the outcome is often expected. Such settings are increasingly studied in the optimal regimes literature \citep{cui2021necessary,cui2021semiparametric,han2021comment,miao2018identifying, qi2021proximal,kallus2021minimax,Cui2021Individualized}. % For example,  \citet{cui2021semiparametric} studied the effect of having a third child on remaining in the labour market among mothers who already had at least two children, using data from \citet{angrist1996children}. The aim of \citet{Cui2021Individualized} was to "provide a personalized recommendation", based on baseline covariates $L$. Beyond using the baseline covariates $L$, however, the $L$-superoptimal regime would further rely on whether the mother \textit{intends} to have three or more children. In the observed data, this intention is deterministically related to actually having three or more children at the census time. Yet, for a future women seeking advice, we could nevertheless imagine asking the question "do you intend to have a third child?"

\section{Superoptimal regimes and their properties}
\label{sec: superoptinal regimes}

Our first proposition states that $L$-superoptimal regimes are always better than, or as good as, $L$-optimal regimes.

\begin{proposition}[Superoptimality]
\label{thm: superoptimality}
The expected potential outcome under the $L$-superoptimal regime is better than or equal to that under the $L$-optimal regime, 
$$
   \mathbb{E}(Y^{g_{\mathbf{opt}}} \mid L = l) \leq  \mathbb{E}(Y^{g_{\mathbf{sup}}} \mid L = l)  \text{ for all } l \in \mathcal{L}.
$$
\end{proposition}
\begin{proof}
Using laws of probability and Definitions \ref{def: opt} and \ref{def: supopt}, 
\begin{align}
      \mathbb{E}(Y^{g_{\mathbf{opt}}} \mid L = l) & = \sum_{a'}  \mathbb{E}(Y^{ g_{\mathbf{opt}}} \mid A=a', L = l) P(A = a' \mid L = l)  \nonumber \\ 
     & \leq \sum_{a'}  \mathbb{E}(Y^{g_{\mathbf{sup}}} \mid A=a', L = l)  P(A = a' \mid  L = l) \nonumber \\ 
      & = \mathbb{E}(Y^{g_{\mathbf{sup}}} \mid L = l)  ,
\end{align}
where the inequality follows because, by definition of $g_{\mathbf{opt}}$ and $g_{\mathbf{sup}}$, we have that $$\mathbb{E}(Y^{ g_{\mathbf{opt}}} \mid A=a', L = l)  \leq  \mathbb{E}(Y^{g_{\mathbf{sup}}} \mid A=a', L = l),$$ for each $a'$.
\end{proof}
Proposition \ref{thm: superoptimality} is not surprising, because the regime $g_{\mathbf{sup}}$ uses more observed information compared to $g_{\mathbf{opt}}$; that is, the $L$-superoptimal regime is optimized not only with respect to $L$ but also with respect to $A$. A similar argument has appeared in \citet{bareinboim2015bandits} for an online bandit setting with no additional covariates, proposing that rewards are maximized when an agent bases decisions on their natural treatment choice.

In the remainder of the manuscript, we will assume that interventions on the treatment variable $A$ are well-defined, such that the following causal consistency assumption holds. 
\begin{assumption}[Consistency] \label{ass: cons}
If $A=a$ then  $Y = Y^a$ for $a\in \{0,1\}$.
%If $A=g(A,L)$ then  $Y = Y^{g}$ for all regimes $g$.

\end{assumption} 

\begin{remark}\label{rem: cons}
Consistency in Assumption \ref{ass: cons} can equivalently be formulated as $Y=Y^A$. This formulation highlights that the factual outcome is equivalent to a particular counterfactual outcome under a regime that assigns treatment $A^{g+}$ according to the trivial regime $g(A,L) = A$ for all patients. Thus, the factual regime is a member of the class of regimes that depends on $A$ and $L$, among which $g_{\mathbf{sup}}$ is the one that maximizes the expected potential outcome. Thus, under consistency, the expected potential outcome under the $L$-superoptimal regime is better than or equal to that under the factual regime.
\end{remark}

We will also invoke the usual positivity assumption.  
\begin{assumption}[Positivity] \label{ass: pos}
$P(A=a \mid L) > 0$ w.p.1 for all $a\in \{0,1\}$.
\end{assumption} 

The following lemma, which exploits positivity and consistency, is similar to arguments that have appeared in work on treatment effects on the treated \citep{robins2007causal,dawid2022can,geneletti2011defining,bareinboim2015bandits}, and will be used in our derivations of identification results. 

\begin{lemma}
\label{lemma: functional}
Under consistency and positivity, $\mathbb{E}(Y^{a} \mid A = a', L=l)$ for $a,a'\in \{0,1\}$ and $l \in \mathcal{L}$ can be expressed as 
\begin{align}
 \mathbb{E}(Y^{a} \mid A = a', L=l)
 & =  \begin{cases}
                    \mathbb{E}(Y \mid A = a', L=l),   & \text{if } a = a' , \\
                    \frac{\mathbb{E}(Y^{a} \mid L=l) - \mathbb{E}(Y \mid A = a, L=l)P(A =a \mid L = l)   }{P(A =a' \mid L = l)},  & \text{if }  a \neq a'. 
                \end{cases} \label{eq: lemma}
\end{align}
\end{lemma}
\begin{proof}
When $ a = a'$, the equation holds by consistency. When $ a \neq a'$, the result follows from the law of total probability, positivity and consistency. 
\end{proof}

Based on Lemma \ref{lemma: functional}, we can use simple algebra to derive the following result, also leveraged by \citet{bareinboim2015bandits} in a setting without covariates. 
\begin{corollary}
\label{cor: sign superopt}
Under consistency and positivity, the $L$-superoptimal regime $g_{\mathbf{sup}}(a',l)$  for $a'\in \{0,1\}$ and $l \in \mathcal{L}$ is equal to
\begin{align}
 g_{\mathbf{sup}}(a',l) = 
 \begin{cases}
                    a'   & \text{if } \ \mathbb{E}(Y \vert L = l) \geq \mathbb{E}(Y^{1-a'}\vert L = l) , \\
                    1-a'  & \text{if } 
 \ \mathbb{E}(Y \vert L = l) < \mathbb{E}(Y^{1-a'}\vert L = l) . 
                \end{cases} \label{eq: sign id exp}
\end{align}
\end{corollary}

The next proposition states conditions for identification of $L$-superoptimal regimes from observed data.

\begin{proposition}[Identification of superoptimal regimes]
\label{thm: id}
Under consistency and positivity, the $L$-superoptimal regime and its value function is identified by the joint distribution of $(L,A,Y)$ whenever
\begin{itemize}
    \item $\mathbb{E}(Y^{a} \mid L=l)$ for all $a \in \{0,1\} $ and $ l \in \mathcal{L}$ is identified. 
\end{itemize}
%If also the distribution of $A$ given $L$ is identified, then the conditional value function under an $L$-superoptimal regime is identified.
\end{proposition}
\begin{proof}
The proposition follows from Lemma \ref{lemma: functional}  and Corollary \ref{cor: sign superopt}, because all the terms on the right hand side of \eqref{eq: lemma} are identified under the two conditions in the proposition. 
\end{proof}

Proposition \ref{thm: id} is useful because it justifies a two-step procedure for identification of $L$-superoptimal regimes using non-experimental data: first, we use (existing) approaches to identify conditional outcome means and the conditional densities of the natural treatment values. Second, we apply the result in Lemma \ref{lemma: functional} to compute counterfactual outcomes conditional on natural treatment values, which allow us to identify $L$-superoptimal regimes. 
Furthermore, Proposition \ref{thm: id} shows that the $L$-superoptimal regime $g_{\mathbf{sup}}$ is identified whenever conditional potential outcomes means, $\mathbb{E}(Y^a \mid L=l) $, are identified in a non-experimental study, which covers studies using IVs or proxy variables as important special cases. 

\begin{remark}[Instrumental variables]
Corollary \ref{thm: id} implies that $L$-superoptimal regimes are identified under assumptions suggested in two recent contributions by  \citet{qiu2021optimal} and \citet{cui2021semiparametric}, who developed theory for identification and estimation of optimal regimes in the presence of unmeasured confounding. That is, under Assumptions \ref{ass: lat unconfoundedness}-\ref{ass: IV Independence} in Appendix \ref{app sec: IVid} the expected outcomes under the regimes given by \citet{qiu2021optimal} and \citet{cui2021semiparametric} will be worse than, or equal to those under the $L$-superoptimal regimes, and, in both cases, the $L$-superoptimal regimes require no extra assumptions for identification of value functions.\footnote{There also exist alternative conditions for identifying optimal treatment rules in IV settings, which only require identification of (the sign of) the causal effect conditional on $L$, and not $\mathbb{E}(Y^{a} \mid L=l)$ itself, as thoroughly discussed by \citet{cui2021necessary}, see also \citet{han2021comment}.}
\end{remark}

\begin{remark}[Proximal inference]
Corollary \ref{thm: id} is also valid in proximal learning settings \citep{miao2018identifying}.
Interestingly, heuristic arguments have been used to justify the inclusion of other covariates, but not the natural value $A$, in the decision function in proximal inference settings. For example, \citet{qi2021proximal} write that  "This may be reasonable since $Z$ may contain some useful information
of $U$, which can help improve the value function."
\end{remark}

We emphasize that the results presented thus far have been agnostic to the absence of unmeasured confounding, which is often equated with the following assumption.

\begin{assumption}[$L$-Exchangeability]
\label{ass: l exch}
$Y^a \independent A  \mid L$ for $a \in \{0,1\}$.
\end{assumption} 

The following results describe different properties of the $L$-superoptimal regime that depend on the truth-value of $L$-Exchangeability. 

\begin{corollary}
\label{cor: equiv}
$L$-Exchangeability implies that $g_{\mathbf{sup}}(A,L) = g_{\mathbf{opt}}(L)$ w.p.1.
\end{corollary}
\begin{proof}
Let $ a^* = \underset{a \in \{0,1\} }{\arg \max } \  \mathbb{E}(Y^{a} \mid  L = l) $.
If $L$-Exchangeability holds, then
$$\mathbb{E}(Y^{a^*} \mid  L = l) = \mathbb{E}(Y^{a^*} \mid A=a',  L = l)   $$ 
for all $a' \in \{0,1\}$ and $l \in \mathcal{L}$. Thus, $a^* = \underset{a \in \{0,1\} }{\arg \max } \  \mathbb{E}(Y^{a} \mid A=a',  L = l) $ for all $a' \in \{0,1\}$.
\end{proof}

\begin{remark}
Suppose that an $L$-superoptimal regime yields better outcomes than an $L$-optimal regime in a given study. Then, it follows from Corollary \ref{cor: equiv} that $L$-Exchangeability fails. This fact can be used to construct tests for unmeasured confounding, see more details in Appendix \ref{appsec: test}. Furthermore, when $L$-Exchangeability fails, an investigator will often assume that there exists a variable $U$, often called an 'unmeasured confounder', that exerts effects on $A$ and $Y$. Then, measuring $U$ in the future will further improve decision making. Because $A$ often represents a decision made by a human in the course of natural practice, then the investigation and measurement of causes of $A$ (e.g., $U$) may be feasible.  

%, then this $U$   would indeed expect that $U$ is observable, at least in principle, because some observed decisions did depend on $U$. 

\end{remark}

\begin{corollary}
\label{cor: equiv2}
Consistency implies that $\mathbb{E}(Y^{g_{\mathbf{sup}}}) \geq \mathbb{E}(Y)$.
When $L$-Exchangeability additionally holds, then $\mathbb{E}(Y^{g_{\mathbf{opt}}}) \geq \mathbb{E}(Y)$.
\end{corollary}
\begin{proof}
As in Remark \ref{rem: cons}, $Y=Y^A$ is generated under a special case of a regime that depends on the natural value of treatment, where $A^{g+} = g(A,L)=A$ w.p.1. Because $g_{\mathbf{sup}}$ is the optimal such regime, then $\mathbb{E}(Y^{g_{\mathbf{sup}}}) \geq \mathbb{E}(Y)$. When $L$-Exchangeability holds, application of Corollary \ref{cor: equiv} completes the proof.
\end{proof}

\begin{remark}
 Given an identified optimal regime, suppose that a human care provider insists that their own intuition about treatment decisions is superior, due to their own access to privileged observations not used by the regime. Corollary \ref{cor: equiv} highlights that this insistence is contradicted when the optimal regime is identified under assumptions of no-unmeasured confounding. Their claim might be depicted by paths in the SWIG of Figure \ref{fig:superoptimal}: if this privileged information was truly useful for decision-making ($U \rightarrow Y^g$) \textit{and} was leveraged by the clinician in the observed data ($U \rightarrow A$), then $Y^a \not\independent A  \mid L$. 
\end{remark}

\begin{remark}
Previous work provides optimality guarantees that exclude $L$-superoptimal regimes. \citet{kallus2021minimax} considered a setting with unmeasured confounding and identified $L$-optimal regimes that are guaranteed to be as good as a ``baseline'' regime. However, this ``baseline'' regime is restricted to be a function of measured baseline covariates $L$. Both the factual regime and the $L$-superoptimal regime are function of unmeasured factors $U$ when there is unmeasured confounding and, thus, does not qualify as a ``baseline'' regime. \citet{ben2021safe} gave a similar safety guarantee, requiring the ``baseline'' regime to be a (deterministic) function of $L$. Thus, these safety guarantees in general neither cover the factual regime nor the superoptimal regime.
\end{remark}

\section{On Experimental Data}\label{sec: experim}
We have thus far only considered observed data $(L,A,Y)$ generated in a non-experimental setting. As anticipated in Definition \ref{def: natval}, we did so because we leverage the natural value of treatment: the treatment an individual would choose in the \textit{absence of it being assigned by an intervention}. In a non-experimental setting, no intervention is made and the treatment a patient actually receives, $A$, is indeed equal to this \textit{natural value}. In experimental settings, however, the patient's natural treatment intentions may be subverted by the experimental design. Therefore, we must introduce additional notation to disambiguate patients' \textit{received} and \textit{intended} treatments in the factual data. Thereby, we let $A^*$ denote the treatment a patient actually receives. Formally, we define a setting to be non-experimental when $A=A^*$ w.p.1, such that the actual treatment value equals the intended treatment value in the setting we have considered so far. In contrast, $A$ may not equal $A^*$ in an experimental setting  Here we discuss several consequences of this distinction, including strategies for identifying the $L$-superoptimal regime with experimental data that differ from those for the non-experimental setting.

A first consequence of the experimental setting is that Assumption \ref{ass: cons} (Consistency), as defined, will almost certainly be violated; in an experiment, a patient actually receives the treatment value corresponding to $A^*$. To illustrate the argument, consider the assumption that $Y=Y^{A^*}$ in an experimental setting. If $A\neq A^*$ for some individuals then $Y=Y^{A^*} \neq Y^A$ for those individuals, thus contradicting Assumption \ref{ass: cons} (Consistency).  Instead the following assumption is more reasonable:    

%we would tend not to assume that a patient's treatment intention $A=a$ will imply equality between their factual outcome and the potential outcome consistent with that value $a$, $Y^a=Y$; the treatment this patient \textit{actually} receives will often differ from that value, $A^*=a' \neq a$.  

\begin{assumption}[Consistency in an experiment] \label{ass: consexp}
If $A^*=a$ then  $Y = Y^a$ for $a\in \{0,1\}$.
\end{assumption}

The results in Section \ref{sec: superoptinal regimes} all suppose Assumption \ref{ass: cons} and not Assumption \ref{ass: consexp}. Thus, they will not in general apply to experimental data. Similar reasoning has historically motivated alternative trial designs, like patient preference trials, in which investigators would attempt to measure $A$ and $A^*$ concurrently, see for example \citet{rucker1989two, knox2019design, forney2019counterfactual}.

%Notably Assumption \ref{ass: consexp} is by definition equivalent to Assumption \ref{ass: cons} when data are non-experimental, so that previous results will apply when premises are adapted to replace Assumption \ref{ass: consexp}  by Assumption \ref{ass: cons}.  

A second consequence of the experimental setting is that $A^*$ is usually allocated such that $Y^a \independent A^*$ by design, and so covariates $L$ will be measured for reasons other than confounding control. Therefore, it is unlikely that $L$-Exchangeability (Assumption \ref{ass: l exch}) will hold in experimental data, as defined. Instead the following assumption is more reasonable:

\begin{assumption}[$L$-Exchangeability in an experiment]
\label{ass: l exch exp}
$Y^a \independent A^*  \mid L$ for $a \in \{0,1\}$.
\end{assumption} 

Despite the irrelevance of the results in Section \ref{sec: superoptinal regimes}, the experimental setting may seem especially appealing for $L$-superoptimal regime identification: because $A^*$ is randomized by design, we can adopt an even more elaborate exchangeability assumption that includes $A$ as a covariate.

\begin{assumption}[$(L, A)$-Exchangeability in an experiment]
\label{ass: l exch exp2}
$Y^a \independent A^*  \mid L, A$ for $a \in \{0,1\}$.
\end{assumption}

Furthermore, we do not have $A=A^*$ by definition, and so the following positivity condition will usually hold:

\begin{assumption}[Positivity in an experiment] \label{ass: pos exp}
$P(A^*=a \mid L, A) > 0$ w.p.1 for $a\in \{0,1\}$.
\end{assumption} 

Lemma \ref{lemma: functional} permitted identification of the $L$-superoptimal regime even when Assumption \ref{ass: pos exp} is contradicted, as it is in a non-experimental setting. With assumptions \ref{ass: consexp}, \ref{ass: l exch exp2}, and \ref{ass: pos exp} and experimental data \citep{forney2019counterfactual}, for example as in patient preference trial designs,  we might trivially identify the $L$-superoptimal regime without appealing to Lemma \ref{lemma: functional}.   In this second approach, the natural treatment value $A$ is simply considered as an additional covariate, and thus effectively subsumed into $L$. 

\begin{lemma} \label{lemma: lemma2} Under consistency, positivity, and $(L,A)$-Exchangeability in an experiment (Assumptions \ref{ass: consexp}, \ref{ass: l exch exp2}, and \ref{ass: pos exp}),

\begin{align}
 \mathbb{E}(Y^{a} \mid A = a', L=l)
 & = \mathbb{E}(Y \mid A^*=a, A = a', L=l). \label{eq: lemma2}
\end{align}
\end{lemma}
\begin{proof}
The equality holds through sequential application of  Assumptions \ref{ass: consexp} and \ref{ass: l exch exp2}, where Assumption \ref{ass: pos exp} ensures that the functional remains well-defined for all values of $a$, $a' \in \{0,1\}$.
\end{proof}

Unfortunately, the natural treatment value $A$ is not measured in most experimental settings. Therefore, when only experimental data are available and $A$ is unmeasured, then Lemma \ref{lemma: lemma2} cannot be used to identify the $L$-superoptimal regime. However, the $L$-optimal regime can be learned with such data via identification of $\mathbb{E}(Y^a \mid L=l)$. The claim and proof is trivial, by considering Lemma \ref{lemma: lemma2} without $A$ in the conditioning set and replacing Assumption \ref{ass: l exch exp2} by Assumption \ref{ass: l exch exp}. 

\begin{remark}
While not useful on its own for learning the $L$-superoptimal regime, knowledge of the parameters $\mathbb{E}(Y^a \mid L=l)$ from an experiment will be instrumental as a supplement to non-experimental data, even if  $L$-Exchangeability (Assumption \ref{ass: l exch}) does not hold for those non-experimental data. Suppose that the non-experimental data and the experimental data are random draws from the same superpopulation. If the conditions of Lemma  \ref{lemma: functional} are met for the non-experimental data, then its identification functional can be evaluated using the combination of the parameters $\mathbb{E}(Y^a \mid L=l)$ learned in the experiment, and those parameters of $(L,A,Y)$ directly observed in the non-experimental setting. This heuristic for combining experimental and non-experimental data has been suggested by \citet{bareinboim2015bandits} for the identification of a $\emptyset$-superoptimal regime. Furthermore, patient preference trials ensure the availability of such data by design, regardless of whether the natural treatment values $A$ are measured in the assigned treatment arms. As an illustration, consider the ICU setting that we introduced in Section \ref{sec: pot out and nat tr}. We could construct a three-arm trial where a doctor first is asked whether they will promptly transfer a patient to an ICU. After recording the answer, we randomly assign the patient to ICU admission, no ICU admission or following the doctor's preference. 
\end{remark}

%This suggested combination of parameters -- learned separately in experimental and non-experimental settings -- is not a novel contribution to the literature. This strategy will be especially convenient when data are available from a 3-arm trial, as described by \citet{knox2019design,long2008causal,forney2019counterfactual}, or when non-experimental data are available from the same population of patients as those sampled for an experiment. 

\section{On algorithm-assisted human decision making}
\label{sec: algorithm ass decision making}

One vision for optimal regimes is to use them in an algorithmic treatment-assignment paradigm, wherein treatments are assigned completely according to learned algorithms without human intervention. This algorithmic paradigm would replace current paradigms centered on consensus standards-of-care guidelines and human care-providers' intuition, which could be fallible. However, the medical community may be resistant to ceding control to such algorithms in the absence of theoretical guarantees that expected outcomes will be better under the targeted optimal regime. We have showed in Corollary \ref{cor: equiv2} that the superiority of the $L$-optimal regime is indeed guaranteed whenever $L$-Exchangeability holds. Nevertheless, the medical community has historically expressed a deep skepticism to $L$-Exchangeability or any of identification strategy that depends on independence conditions in non-experimental data; see for example the Journal of the American Medical Association's prohibition on causal language for the results of non-experimental studies \citep{ama2020ama}. When an $L$-optimal regime is learned in the absence of $L$-Exchangeability --- for example, when the $L$-optimal regime is learned using data from a conventional 2-arm trail --- a clinician's skepticism may be justified: we cannot guarantee that $$\mathbb{E}(Y^{g_{\mathbf{opt}}} ) \geq \mathbb{E}(Y^{A}).$$ 

A primary benefit of the super-optimal regime $g_{\mathbf{sup}}$ is to provide an algorithm with guarantees that $$\mathbb{E}(Y^{g_{\mathbf{sup}}}) \geq \mathbb{E}(Y^{A}).$$ We illustrated in Section \ref{sec: experim} that the super-optimal regime is estimable from a combination of experimental and non-experimental data whereby all relevant assumptions are enforced by design; thus such results may be acceptable to a skeptical medical community. 

Nevertheless, current formulations of the $L$-superoptimal regime $g_{\mathbf{sup}}$ consider treatment intentions $A$ as simply an additional covariate. Thus, this formulation suggests a paradigm where the algorithm is rhetorically centered. This radical departure from existing treatment assignment paradigms may result in the persistence of skepticism and resistance, despite the guarantees of the $L$-superoptimal regime. Therefore, we provide the following equivalent formulation of $g_{\mathbf{sup}}$ that suggests a paradigm where the human care provider remains centered.

\begin{proposition}
    \label{prop: reform}

There exists a function $\gamma: \mathcal{L} \rightarrow \{0,1,2\}$ such that the following equality holds w.p.1,
\begin{align}
   g_{\mathbf{sup}}(A,L)  = \begin{cases}
                                g_{\mathbf{opt}}(L) &\text{ if  }\ \  \gamma(L) = 0 \\
                                A &\text{ if  }\ \   \gamma(L) = 1 \\
                               1-A &\text{ if  } \ \  \gamma(L) = 2.
                            \end{cases}
                             \label{eq: reform} 
\end{align}

\end{proposition} 

The function $\gamma(l)$ is identified as 
\begin{align*}
\gamma(l) = 
    \begin{cases}
        0 & \text{ if } \{\tau_l(1)\geq0, \tau_l(0)\geq0\} \text{ or } \{\tau_l(1)<0, \tau_l(0)<0\} \\ 
        1 & \text{ if } \{\tau_l(1)\geq0, \tau_l(0)<0\} \\
        2 & \text{ if } \{\tau_l(1)<0, \tau_l(0)\geq0\},
    \end{cases}
\end{align*}
where we let $\tau_l(a') \coloneqq \mathbb{E}(Y^{a=1}  \mid A=a', L=l) - \mathbb{E}(Y^{a=0}  \mid A=a', L=l)$.
A proof is provided in Appendix \ref{appsec: Prop3proof}. Proposition \ref{prop: reform} formulates an algorithm that directly negotiates between the $L$-optimal regime $g_{\mathbf{opt}}$, and a human care provider's own privileged intuition, captured by their natural treatment intention $A$: when a provider encounters a patient, they are given the value of the random variable $\gamma(L)$; if $\gamma(L) = 0$, then the provider is instructed to follow the $L$-optimal regime's recommendation, $g_{\mathbf{opt}}(L)$; if $\gamma(L) = 1$ then the provider is instructed to override the $L$-optimal regime's recommendation and provide the treatment according to their natural intention, $A$.  Finally, if $\gamma(L) = 2$ then the provider is instructed to override the $L$-optimal regime's recommendation and provide the treatment \textit{opposite} to their natural intention, $1-A$. With this formulation, superoptimal regime methodology can be described as a strategy for optimally negotiating between a typical $L$-optimal regime and a provider's privileged intuition: when the $L$-optimal regime is already known, the function $\gamma$ can be learned to indicate to a care provider when the $L$-optimal regime should be followed, or else should be overridden as a function of their natural treatment intention $A$. Because this formulation is equivalent to the $L$-superoptimal regime $g_{\mathbf{sup}}$, then the provider has guarantees that this algorithm will outperform the status quo. Thus, the use of $L$-superoptimal regimes is accurately described as ``algorithm-assisted human decision making''.

However, the term ``algorithm-assisted human decision making'' has also been used to describe settings where the decision maker receives information from an algorithm and subsequently makes their decision. For example, consider an experiment that randomly assigned judges to receive no information or output from a Public Safety Assessment algorithm \citep{imai2023experimental}. The algorithmic output included a recommended decision and the predicted risks of certain adverse outcomes. The judges made their autonomous decisions after receiving this information. The motivation for providing the algorithmic output matches the motivation for $L$-optimal regimes, at least when adhering to classical decision making criteria \citep{sarvet2023perspectives,sarvet2023aaron,stensrud2023discussion}: that is, finding a decision rule ``that minimizes the prevalence of negative outcomes while avoiding unnecessarily harsh decisions''\citep{imai2020principal}. 
However, this type of algorithm-assisted decision making could easily be augmented with algorithmic output of $L$-superoptimal regimes. Specifically, the algorithm could output $g_{\mathbf{sup}}(a',L)$ for both $a' \in \{0,1\}$. The decision maker could receive this information before they state their intended treatment value; the $L$-superoptimal regimes are provided for each intended treatment value. Indeed, the decision maker can use the algorithmic output even without uncovering their intended treatment $A$. For example, an algorithm might inform an ICU doctor that for a patient with covariates $l$ the recommended treatment is  $g_{\mathbf{sup}}(a',l)=a'$ for both $a' \in \{0,1\}$. This means that the algorithm supports the doctor's intended decision for such a patient, whatever it may be. So, if an ICU doctor plans to give treatment $a'=1$, they receive confirmation that this aligns with the algorithm's recommendation. However, if the algorithm suggests $g_{\mathbf{sup}}(a',l)=a$ for any $a'\neq a$, then a  doctor who intended to give such a treatment $a'$ would be alerted to a discrepancy with the algorithm's recommendation. They then have the option to reconsider their decision. It is possible that $g_{\mathbf{opt}}(l)=a'$ in both scenarios, but only the second scenario ensures that $g_{\mathbf{opt}}(l)=g_{\mathbf{sup}}(a',l)=a'$ for $a'\in \{0,1\}$. Therefore, knowing $g_{\mathbf{sup}}(a',L)$ for $a' \in \{0,1\}$ gives the ICU doctor additional information to support their decisions, even without the doctor disclosing their intentions. In this way, while the $L$-superoptimal regimes ostensibly reverse the order of the algorithm and the human in the decision making, we can unify these roles.

%Although the recommended decisions and risk scores are not guaranteed to correspond to $g_{\mathbf{opt}}(L)$, or value functions under $g_{\mathbf{opt}}(L)$, the algorithmic choice of algorithmic output seems to have an analogous motivation; pretreatment covariates $L$ are used to make an individualized decision that

We emphasize that any override of the original algorithm, beit $g_{\mathbf{opt}}$ or $g_{\mathbf{sup}}$, will in general forgo the optimality guarantees of that original algorithm. In particular, we cannot formally guarantee that the decision maker, based on their free will, would make better decisions using algorithmic information. To study such decision settings, we would need different (experimental) data, where a decision maker's intended treatments are measured both before and after they are provided algorithmic information.

%%A similar concern is relevant to the generalizability of $L$-superoptimal regimes: in a future decision setting, individuals might not make intended decisions ($A$) in the same way as individuals did under the observed data law. 
The effect of algorithms on natural human decisions themselves is related to a more general complication that may arise with the use of $L$-superoptimal regimes in practice \citep{mclaughlin2022algorithmic}. While an investigator might expect the conditional distribution of the potential outcomes given covariates,  $f_{Y^a\mid L}$, to remain stable across time, e.g., for biological reasons, they might not expect the same stability in the conditional distribution of the natural treatment given covariates $f_{A\mid L}$. For example, this conditional distribution may change when individuals know that an algorithm will use $A$ as input. When this generalizability problem is present, we  cannot guarantee the performance of $L$-superoptimal regimes in the future decision setting, as the $L$-super-optimal regime is a function of $f_{A\mid L}$.  In contrast, we might still have guarantees for $L$-optimal regime, because this regime is not a function of $f_{A\mid L}$. However, to guarantee that $g_{\mathbf{opt}}(L)$ in the observed data setting corresponds to the $L$-optimal regime in a future decision setting, we still need to make assumptions about shared probabilistic structure across these settings, see e.g.\  \citet{bareinboim2013general} and \citet{dahabreh2019generalizing}. Thus, the conventional $L$-optimal regimes also require strong assumptions for generalizability. To have analogous guarantees for $g_{\mathbf{sup}}$, we would further need to make assumptions on the probabilistic structure of the natural value. For example, it would be sufficient, but not necessary, that individuals make their intended decision $A$ in the same way they would in the observed data. The plausibility of this condition is context-dependent. Suppose that a decision maker receives the results of a study showing that the superoptimal regime agrees with the intended (natural) treatment value for individuals with a particular covariate value $l$. Then, we would not expect the decision maker to change their natural decisions for such individuals in a future decision setting; the study just confirmed that their intended treatment in this context is the best treatment option, given the observed data, and we might believe that $f_{A\mid L}$ is stable. However, suppose instead that a decision maker receives the results from a study showing that their natural decisions for individuals with covariate value $l$ are opposite of those recommended by the superoptimal regime. If the decision maker changes their natural decision process after seeing these study results, then we would not expect that $f_{A\mid L}$ is stable. To mitigate the problem of an unstable distribution, we could, in the future decision setting, attempt to retrieve an intended treatment value that is representative of the study result. For example, we could instruct the decision maker to provide the previously-learned algorithm their intended treatment \textit{had} they not received the recent study results. We leave to future work the elaboration of weaker conditions for the stability of super-optimal regimes. 

%However, in certain settings individuals might change their own decisions when knowing that they might, eventually, be overruled by an algorithm. Furthermore, individuals might change their decisions in the future because they have gained new knowledge. 

%This is \citep{mclaughlin2022algorithmic} discussed the fact that decision makers might change their opinions when they know the algorithmic output. However, we consider a setting where the individuals need to make an intended decision before seeing the algorithmic recommendation. 

%Our $L$=superoptimal regimes ostensibly reverse the order of the algorithm and the human, compared to the setting e.g.\ discussed in \citep{mclaughlin2022algorithmic}. However, we could also let the $L$-superoptimal regimes be given to the decision maker, who then has the final saying. When we let the decision maker have the final saying, we lose the formal guarantees of optimaliyy compared to the $L$-superoptimal regime; it might be that the decision maker makes better or worse decisions. However, compared to the current recommendation systems, we could give then $L$-superoptimal regimes for either value of $A$. Indeed, from basic decision theory, we would expect that this guarantees the humans to do even better decisions (conjecture). Indeed, when the pair of $L$-superoptimal regimes agree, we know they are equal to the $L$-optimal regime. If the pair of $L$-superoptimal disagree, i.e. have different signs, then the decision maker should interpret this as being a setting where overruling the $L$-optimal regime actually matters.

Finally, consider a future decision setting where both the $L$-optimal and $L$-superoptimal regime maintain their nominal guarantees. Like classical settings without unmeasured confounding, the \textit{estimated} $L$-optimal and $L$-superoptimal regime, learned with finite data, might differ from the \textit{true}  $L$-optimal and $L$-superoptimal regime \citep{hubbard2016statistical}. Thus, due to sampling variability, the \textit{estimated} $L$-superoptimal and $L$-optimal regimes might perform worse than the \textit{true} $L$-superoptimal and $L$-optimal regimes, respectively. Similarly, the \textit{estimated} $L$-superoptimal regime might also perform worse than the natural (observed) regime. In future work, we will study strategies to control the error of deviating from the natural regime when the natural regime actually is optimal, based on family wise error rates and false discovery rates.  %This is not a unique problem for the $L$-superoptimal regime.

\section{On the non-prescriptive use of superoptimal regimes}
\label{sec: alternative use}

The formulation of  $g_{\mathbf{sup}}$ in Proposition \ref{prop: reform} highlights a counterintuitive possibility of an $L$-superoptimal regime: when $P(\gamma(L)=2)>0$, the  $L$-superoptimal regime indicates that a decision maker should assign precisely the treatment value that is \textit{the opposite} of their natural intentions, $1-A$, for some patients. This could be the case when humans currently use outcome-predicting variables in precisely the \textit{opposite} way from that which would optimize outcomes.\footnote{Analogous to our discussion in Section \ref{sec: algorithm ass decision making}, we caution that the estimated superoptimal regime $\hat g_{\mathbf{sup}}$ might differ from true $g_{\mathbf{sup}}$ due to sampling variability.}

An algorithm-driven health care system might dismiss this occurrence as an ancillary curiosity; if $\gamma(L)=2$, then providing treatment $1-A$ would simply be the optimal choice, given covariates $A$ and $L$. However, $g_{\mathbf{sup}}$ is more than simply a prescriptive treatment policy; a positive probability of $\gamma(L)=2$ might indicate an opportunity to radically adjust existing theories or systems for patient care for some groups, which were apparently grossly misformulated. 
The history of the study of human behavior offers many examples of fallacies where humans systematically (but unintentionally) undermine their own objectives, and iatrogenic harm is one well-documented subclass of this phenomenon. Detecting these occurrences is surely an important scientific aim, as major paradigm shifts in medical history have been portended by the scientific communities attention to such paradoxes \citep{kuhn1970structure}.

%We briefly review one such paradigm shift, now regarded as having portended a revolution in medical hygiene decades before germ theory. 

\begin{example}[Semmelweis]
Consider the case of Ignaz Semmelweis, a 19th-century Hungarian physician. Semmelweis famously observed that it was precisely the women who were admitted to elite teaching hospital wards in anticipation of obstetric complications $(A=1)$ who were experiencing increased mortality from puerperal fever. This was an ostensibly paradoxical observation: the elite venues ($A=1$) purported to offer the best possible care. If Semmelweis had used data to learn the $L$-superoptimal regime, he would have observed that $P(\gamma(L) = 2)>0$; that is, there exist subgroups where the best thing to do is to \textit{not} admit to the elite teaching hospital ($A^{g_{\mathbf{sup}}+}=0$) precisely those patients who would otherwise be admitted to such a ward ($A=1$), and admit those patients ($A^{g_{\mathbf{sup}}+}=1$) who would otherwise be treated in a less prestigious venue ($A=0$). Semmelweis ultimately uncovered an explanation: women sent to the prestigious hospitals were the most likely to need surgical intervention, which was then often provided by physicians returning from autopsy procedures with hands unwashed \citep{semmelweis1983etiology}.  Semmelweis's observations helped initiate a hygiene and hand-washing revolution in medicine. 

Semmelweis did not need the formalisms of super-optimal regimes to make his discovery. Instead, he relied on savvy intuition and large effect sizes. Superoptimal regime methodology provides a tool for systematic surveillance of (iatrogenic) harm, even when effect sizes are modest, or human intuition would otherwise fail.
\end{example}

\section{Case study: Instrumental variables}
\label{sec: case study iv}

The results we have derived so far are general. They can be used in any setting where value functions and the joint distribution of the factuals $(L,A,Y)$ are identified. Thus, these results could be of interest in a range of settings where investigators would otherwise aim to find $L$-optimal regimes in the presence of unmeasured confounding. In each particular setting, an investigator can derive explicit identification formulae, which in turn motivate estimators.

In our first case study, we re-visit an example from the seminal paper by \citet{BalkePearl1997}, illustrating that $L$-superoptimal regimes for certain values of $A$ can be point identified even if $L$-optimal regimes are not. 

\begin{example}[Vitamin A supplementation and mortality] 
 \citet{BalkePearl1997} derived bounds for average causal effects in instrumental variable settings. In particular, these bounds are sharp under an individual level exclusion restriction, $Y^{a,z}=Y^{a} $ for all $ a,z,$ and the exchangeability assumption $ Z \independent \{ Y^{a=1}, Y^{a=0} , A^{z=0}, A^{z=1}\}$, which holds by design when $Z$ is randomly assigned. The sharpness guarantees also hold under weaker assumptions \citep{swanson2018partial}. To illustrate the practical relevance of these bounds, \citet{BalkePearl1997} analyzed data from a randomized experiment in Northern Sumatra, where 450 villages were randomly offered oral doses of Vitamin A supplementation ($Z=1$) or no treatment ($Z=0$). Villages receiving Vitamin A supplementation were encouraged to provide them to pre-school children of age 12-71 months. The dataset contained 10231 individuals from villages assigned to Vitamin A ($Z=1$) and 10919 untreated individuals ($Z=0$). \citet{BalkePearl1997} studied the effect of consuming Vitamin A supplementation ($A=1$) vs no treatment ($A=0$) on survival ($Y=1$) after 12 months. Leveraging that $Z$ is an instrument, \citet{BalkePearl1997} reported bounds on the average treatment effect,
$$
-0.1946  \leq \mathbb{E}(Y^{a=1} - Y^{a=0}) \leq  0.0054.
$$

They concluded that the "Vitamin  A  supplement,  if  uniformly administered, is seen as capable  of increasing  mortality  rate  by much as 19.46\% and is incapable  of reducing mortality rate by more than 5.4\%". \citet{BalkePearl1997} did not consider further covariates, and thus we define $L = \emptyset$. It follows that neither the $L$-optimal regime, $$ {g_{\mathbf{opt}}}  = \underset{a \in \{0,1\} }{\arg \max } \ \mathbb{E}({Y}^a),$$ nor the value function $\mathbb{E}({Y}^a)$ are point identified, but both functionals are non-trivially bounded.  However, consider now the regime $${g_{\mathbf{sup}}} (a')  = \underset{a \in \{0,1\} }{\arg \max } \ \mathbb{E}(Y^{a} \vert A = a'),$$ which uses the intended value of Vitamin A consumption as input to the decision function. Using Lemma \ref{lemma: functional} in this particular example, we have point identification of the superoptimal regimes for the treated, which is an analogous parameter to the ATT,
$$\mathbb{E}(Y^{a = 1} - Y^{a = 0} \mid A = 1) = 0.0032.
    $$
This example illustrates the point that, under conditions that do not identify an $L$-optimal regime recommendation, the $L$-superoptimal regime recommendation for a particular value of $A$ is identifiable. Indeed, we conclude that among children who would consume Vitamin A supplementation when offered, the Vitamin A supplementation does have a beneficial effect.\footnote{In this example, the point identification of the superoptimal CATE follows because Vitamin $A$ treatment was inaccessible to those randomly assigned to no treatment ($Z=0$), see \citet{BalkePearl1997}[Table 1], that is, there is one-sided compliance.} 

Whereas the $L$-superoptimal regime recommendation given $A=1$ is point identified, the $L$-superoptimal regime recommendation given $A=0$ is not, that is,
$$
-0.33 \leq \mathbb{E}(Y^{a = 1} - Y^{a = 0} \mid A = 0) \leq 0.0069.
$$
\end{example}

\subsection{Point identification of value functions}

To illustrate how explicit identification formulae and estimators can be derived,  we further build on recent work on optimal regimes \citep{qiu2021optimal,Cui2021Individualized,cui2021necessary}. These results are given in detail in the online appendices, and we provide an overview in this section. Specifically, we give new identification results for $L$-superoptimal regimes (Appendix \ref{app sec: IVid}). We derive the non-parametric influence function of corresponding identification functionals (Appendix \ref{appsec: influence function id}), and thereby motivate new estimators (Appendix \ref{appsec: influence function id} and \ref{appsec: estim algorithm}). Furthermore, we suggest a strategy to further improve efficiency when the investigator only aims to identify the $L$-superoptimal regime (given by the sign of the value function, see Corollary \ref{cor: sign superopt}), as opposed to the value function itself.
We apply these new methods to study the effects of ICU admission on survival in Section \ref{sec: icu example}. %To motivate these developments, we first consider a synthetic example from \citet{qiu2021optimal}.

To illustrate the practical benefits of superoptimal vs optimal regimes in this setting, we revisit in Appendix \ref{appsec: ex} an example from \citet[Remark 5]{qiu2021optimal}, who emphasized that an $L$-optimal regime can be worse than the regime that was implemented in the observed data. We show that the $L$-superoptimal regime is strictly better than the $L$-optimal regime in their example. We further give another example (See Example \ref{ex: luedkte modif}), with a minor change to the setting in \citet{qiu2021optimal}, where the $L$-superoptimal regime outperforms both the $L$-optimal and the observed regime.

%Appendix \ref{app sec: id iv setting} reviews a sufficient set of assumptions (see \ref{ass: lat unconfoundedness}-\ref{ass: IV Independence}) for identification of average treatment effects using IVs \citep{wang2017identification,cui2021semiparametric,cui2021necessary,qiu2021optimal,angrist1996identification}. We will use this identification result to derive efficient estimators, based on the nonparametric influence function in Appendix  \ref{app sec: influence function id} and \ref{app sec: one step estimator}.

\subsection{Augmenting regimes with instruments}

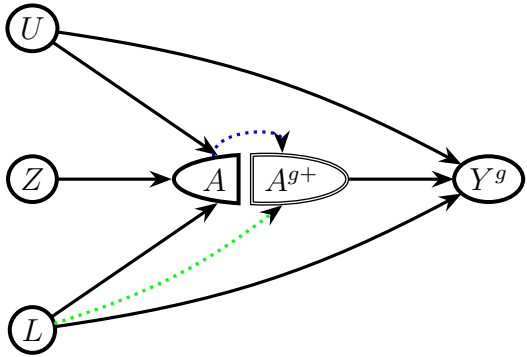
\begin{figure}
            \centering
            \begin{tikzpicture}
                \tikzset{line width=1.5pt, outer sep=0pt,
                ell/.style={draw,fill=white, inner sep=2pt,
                line width=1.5pt},
                swig vsplit={gap=5pt,
                inner line width right=0.5pt},
                swig hsplit={gap=5pt}
                };
                      \node[name=L,ell,shape=ellipse] at (3,-2){$L$};
                    \node[name=A,shape=swig vsplit] at (6,0) {
                                                      \nodepart{left}{$A$}
                                                      \nodepart{right}{$A^{g+}$} };
                    \node[name=Y, ell, shape=ellipse] at (9,0){$Y^{g}$};
                    \node[name=U, ell, shape=ellipse] at (3,2){$U$};
                    \node[name=Z, ell, shape=ellipse] at (3,0){$Z$};
                \begin{scope}[>={Stealth[black]},
                  every node/.style={fill=white,circle},
                  every edge/.style={draw=black,very thick}]
                    \path[->] (U) edge[bend left=10] (Y);
                    \path[->] (U) edge (A);
                    \path[->] (L) edge (A);
                    \path[->] (Z) edge (A);
                    \path[->] (A) edge (Y);
                    \path [->] (L) edge[color=green,dotted, bend right=10] (A.300);
                    \path [->] (L) edge[bend right=10] (Y);
                    \path [->] (A.150) edge[color=blue,dotted,  out=120, in=130, bend left = 90] (A.60);
                \end{scope}
            \end{tikzpicture}
            \caption{A dynamic SWIG with instrumental variable $Z$ describing a regime that depends on $A$ and $L$, consistent with a  superoptimal regime.}
                            \label{fig:superoptimal}
        \end{figure}

There is an important role of instruments in $L$-superoptimal regimes, which differ from their role in $L$-optimal regimes. Let an $(L,Z)$-optimal regime be defined as  
$$ g_{\mathbf{opt}}(l,z)  \equiv \underset{a \in \{0,1\} }{\arg \max } \  \mathbb{E}(Y^a \mid L = l, Z=z),$$
which is isomorphic to an $L$-optimal regime but further uses the instrument $Z$. 
An instrument $Z$ satisfies $Y^a \independent Z  \mid L$, which can be read off of the SWIG in Figure \ref{fig:superoptimal}. Using arguments isomorphic to those of Corollary \ref{cor: equiv}, the $(L,Z)$-optimal regime is always equal to the $L$-optimal regime. However, interestingly, an $(L,Z)$-superoptimal regime is not necessarily equal to an $L$-superoptimal regime. This follows because in many cases
$$
Y^a \not\!\perp\!\!\!\perp Z \mid L,A,
$$
see the SWIG in Figure \ref{fig:superoptimal} as an example. Therein, we see that $Z$ would be d-connected to $Y^a$ given $L$ and $A$ via the path $Z\rightarrow A \leftarrow U \rightarrow Y^a$, which would be open conditional on $A$, a collider. A similarly open path would remain if $Z$ was alternatively associated with $A$ via an unmeasured common cause, which is often assumed in some IV models. The practical implication is that using an instrument $Z$ can further improve superoptimal, but not optimal, regimes, see Appendix \ref{app sec: IVsuperopt} for more details. We give intuition to this result in our ICU example in Section \ref{sec: icu example}. 

%Consider our ICU example. Using the doctor's intended treatment can improve decisions. However, using the doctor's intended treatment 

\section{Application: Intensive Care Unit Admissions}
\label{sec: icu example}

Following \citet{keele2020stronger}, we study the effect of prompt ICU admission on 7-day survival. We use resampled data from a cohort study of patients with deteriorating health who were referred for assessment for ICU admission in 48 UK National Health
Service (NHS) hospitals in 2010-2011 \citep{harris2015delay}. 

Our treatment of interest, $A=1$, is ICU admission within 4 hours upon arrival in the hospital ('prompt ICU admission'). An individual is untreated, $A=0$, if they were not admitted within 4 hours. Our sample consists of 13011 patients, of whom 10478  were treated. One reason for being untreated could be resource constraints, e.g., the lack of available ICU beds or insufficient staffing. Like \citet{keele2020stronger}, we use an indicator of the ICU bed occupancy being below or above the median (4 beds) as our instrument, $Z$, which should only affect the outcome $Y$ through its effect on $A$ (Figure \ref{fig:superoptimal}). We further considered an individual's age, recorded sex and sequential organ failure assessment score as baseline variables ($L$). % sepsis diagnosis, peri-arrest diagnosis, INCARC physiological score, NHS national early warning score, 

In these non-experimental data, the individual's natural value of treatment is directly recorded. In a future decision setting, we could measure the natural treatment variable by asking the following question to a doctor treating a patient: ``Would you promptly admit this patient to an ICU?'' A ``yes'' to this answer would correspond to $A=1$ and a ``no'' would correspond to $A=0$.\footnote{We suppose a deterministic relation between the doctor's response to this question and what they actually would have done.} Informally, the doctor's response, $A$, serves as a proxy for factors $U$ that might indicate the risk of 7-day mortality. Furthermore, using the current bed occupancy $Z$ jointly with the doctor's response $A$ could give a better proxy for factors $U$ not recorded in the observed data, even when ICU bed occupancy $Z$ is independent of $U$ marginally.  For example, suppose $U$ represents a physician's judgement of a patient's underlying mortality risk based on unrecorded injury features or other implicit judgments of patient frailty  (``moderate'' or  ``severe'') and that a doctor will admit all patients on low-occupancy days but will only admit ``severe''-risk patients on high-occupancy days. Occupancy has little predictive capacity for a patient's underlying mortality risk marginally, but if it is known that a patient was admitted on a high occupancy day, then we can deduce that the patient must have been at ``severe'' risk. %Similar logic underlies associations induced by collider stratification in Bayesian networks representing structural causal models.

We estimated observed, $L$-optimal, $L$-superoptimal and  $(L,Z)$-superoptimal regimes based on the estimation algorithm in Appendix \ref{appsec: estim algorithm}, where we also used 60-40 sample splitting to avoid the (positive) bias that would result from estimating and evaluating a (super)optimal decision rule in the same sample \citep{zhang2012robust,qiu2021optimal}. The point estimates of the marginal value functions suggest that the $L$-superoptimal ($\hat g_{\mathbf{sup}}$)  and $(L,Z)$-superoptimal regimes ($\hat g_{\mathbf{z-sup}}$) outperform the alternatives (Table \ref{tab:my_label}). The fact that the confidence intervals are wide is not surprising, despite the large sample size, because of the reliance on an instrumental variable. However, the imprecision also requires us to caution against making strong conclusions about the estimated (super)optimal regimes, because they might deviate from the true (super)optimal regimes due to finite-sample uncertainty, as discussed in Section \ref{sec: algorithm ass decision making}.

\begin{table}[]
    \centering
    \begin{tabular}{c|c}
         Parameter & Estimate (95 \% confidence interval)  \\
         \hline
         $\mathbb{E}(Y)$ & 0.86 (0.85, 0.86) \\
         $\mathbb{E}(Y^{\hat g_{\mathbf{opt}}})$ & 0.93 (0.40, 1.00) \\
         $\mathbb{E}(Y^{\hat g_{\mathbf{sup}}})$ & 0.97 (0.77, 1.00) \\
        $\mathbb{E}(Y^{\hat g_{\mathbf{z-sup}}})$ & 0.98 (0.77, 1.00)
   %      $\mathbb{E}(Y^{g_{\mathbf{opt}}})$ & 0.894 (0.549, 1.392) \\
    \end{tabular}
    \caption{Marginal value functions under different regimes, where the percentile 95\%-confidence intervals are estimated by non-parametric Bootstrap in 500 samples.}
    \label{tab:my_label}
\end{table}

\section{Future directions}
\label{sec: discussion}
An interesting problem is generalizing the results to longitudinal settings with time-varying treatments. A complicating factor is that the non-baseline natural treatment values will not in general correspond with a patient's observed treatment values, even when the data arise from a non-experimental setting.   Nevertheless, their distributions may be identified under assumptions commonly invoked to identify dynamic regimes that depend on the natural value of treatment in time-varying treatment settings, as described in \citet{richardson2013single} and \citet{young2014identification}. Generalizations to non-binary treatments will also be of interest in some settings.

Our identification results motivate estimators of $L$-superoptimal regimes. We specifically gave semi-parametric estimators in an IV setting. However, alternative estimators of superoptimal regimes can also be developed, and the properties of these estimators must be evaluated on a case-by-case basis. Relatedly, we aim to construct estimators of superoptimal regimes with error control, e.g., of erroneously deviating from the observed regime.  %, e.g.\ depending on the identification formula, and estimator, for the conditional mean  $\mathbb{E}(Y^a \mid A=a', L=l) $. As an interesting special case, we derived  

Finally, there exist results on $L$-optimal regime identification when conditional outcome means are only partially identified \citep{pu2021estimating,cui2021machine,Cui2021Individualized}. The partial identification results can e.g.\ be derived under conditions that are guaranteed by design. In contrast, point identification in settings with unmeasured confounding, for example in the IV setting we considered here, requires homogeneity assumptions that are not guaranteed to hold. Using $L$-superoptimal regimes under partial identification conditions is a topic for future investigations.

\subsection*{Acknowledgements}
We would like to thank Prof.\ Vanessa Didelez for insightful comments on a previous version of this manuscript. The data were kindly provided by Prof.\ Luke Keele. 

\bibliographystyle{plainnat}
\bibliography{references}
\appendix

\appendix

\addcontentsline{toc}{section}{Online Appendix}
\section*{Online Appendix}
\renewcommand{\thesubsection}{\Alph{subsection}}
%Short introduction to overall appendix goes here
In this appendix, we present some new results and review existing conditions for identification and estimation of regimes in instrumental variable (IV) settings. 

More specifically, in Section \ref{appsec: IV} we consider identification conditions in IV settings and we describe the implications of these conditions for effect heterogeneity. Furthermore, we make two practical contributions. First, we illustrate that an assumption commonly made in service of IV identification implies that the optimal regime is equivalent to the superoptimal regime, which calls into question its utility in these investigations. Second, we show that, whereas the use of instruments in decision rules is uninformative for conventional optimal regimes, their use improves superoptimal regimes.  In Section \ref{appsec: ex}, we give explicit examples, motivated by the existing literature, wherein  superoptimal regimes outperform optimal regimes in IV settings. In Section \ref{appsec: test}, we use superoptimal regime theory to motivate a novel hypothesis test for detection of unmeasured confounding, leveraging restrictions on the observed data law that follow from the superoptimality property of Proposition \ref{thm: superoptimality}. %where the idea is to assess whether a superoptimal regime outperforms an optimal regime. 
In Section \ref{appsec: est} we describe estimators of superoptimal regimes in IV settings. These estimators are motivated by -- and derived from -- classical results on semi-parametric theory \citep{newey1994, Van2000}. In particular, we describe a multiply robust estimator and also the simple semi-parametric estimator that is used in our data example. In Section \ref{appsec: proofs}, we provide proofs of our formal results, including those of Section \ref{appsec: est}; while the theory therein is a straightforward extension of existing work, for completeness and convenience to the reader, we provide comprehensive arguments in terms of our notation.

%Some of these proofs follow from classical results, but we have nevertheless given comprehensive arguments to make the text self contained. 

\subsection{Instrumental variables (IVs)} \label{appsec: IV}
%Short introduction to IV appendix goes here

\subsubsection{Identification of the superoptimal regime} \label{app sec: IVid}
Here we review sufficient assumptions for identification of $\mathbb{E}(Y^{a} \mid  L = l)$ in IV settings \citep{cui2021necessary}. These assumptions are invoked in Sections \ref{sec: case study iv} and \ref{sec: icu example} of the main text.  To simplify notation, let $f( z \mid l) \coloneqq P(Z=z \mid L=l)$. Furthermore, let  $\delta(l) \coloneqq P(A=1  \mid  Z=1,l)-P(A=1 \mid  Z=0,l)$ and $\widetilde \delta(l,u) \coloneqq P(A=1 \mid Z=1,L=l,U=u)-P(A=1 \mid Z=0,L=l,U=u)$.

\begin{assumption}\emph{(Latent unconfoundedness)}
$Y^{z,a} \independent (Z, A) \mid L, U$ for $z,a \in \{ 0,1 \}$.
\label{ass: lat unconfoundedness}
\end{assumption}
Assumption \ref{ass: lat unconfoundedness} states that adjusting for $U$ and $L$ would, in principle, be sufficient to adjust for confounding between $A$ and $Y$. We further impose the following four IV assumptions, which are standard in the literature, see e.g.\ \citet{cui2021necessary, angrist1996identification}:

\begin{assumption}\emph{(IV relevance)} $Z \not\!\perp\!\!\!\perp A \mid L$.
\label{ass: IV Relevance}
\end{assumption}

\begin{assumption}\emph{(Exclusion restriction)} $Y^{z,a}=Y^a$  w.p.1. for $z,a\in \{ 0,1 \}$.
\label{ass: Exclusion Restriction}
\end{assumption}

\begin{assumption}\emph{(IV independence)} $Z \independent  U  \mid L$.
\label{ass: IV Independence}
\end{assumption}

\begin{assumption}\emph{(IV positivity)} $0<f\left( 1 \mid L\right)<1$ w.p.1.
\label{ass: IV positivity}
\end{assumption}
Finally, we consider the following two assumptions about effect homogeneity  \citep{cui2021necessary,cui2021semiparametric}.
\begin{assumption}\emph{(Treatment homogeneity)} 
\label{ass: indep compliance type}
$ \delta(L) =\widetilde \delta(L,U)~ \text{w.p.1}.$
\end{assumption}

\begin{assumption}\emph{(Outcome homogeneity)} 
\label{ass: indep response type}
$ \mathbb{E} ( Y^{a=1}-Y^{a=0} \mid L) = \mathbb{E} ( Y^{a=1}-Y^{a=0} \mid L,U)~ \text{w.p.1}.$ 
\end{assumption}

As we subsequently show, assumptions \ref{ass: lat unconfoundedness}-\ref{ass: indep compliance type} are sufficient for identifying the super-optimal regime. Importantly these assumptions do not \textit{a priori} preclude $U$ from being a qualitative modifier of the effect of $A$ on $Y$ given $L$, as illustrated by Example \ref{ex: iv heterogeneity} in Appendix \ref{appsec: ex}. Such a preclusion would imply that the value of the $L$-superoptimal regime is identical to that of the $L$-optimal regime. Indeed, another common identification strategy would substitute Assumption \ref{ass: indep response type} for Assumption \ref{ass: indep compliance type}, based on seminal work on structural nested models \citep{robins1994correcting}. However, Assumption \ref{ass: indep response type} implies an equality between $g_{\mathbf{opt}}$ and $g_{\mathbf{sup}}$. If interest lies specifically in differences between the superoptimal vs. the optimal regime in a given setting, an investigator making Assumption \ref{ass: indep response type} based on subject matter knowledge would not need to conduct further analysis of the data.
%\footnote{\citet{cui2021necessary} later demonstrated necessary conditions for identification of sign, CATEs and value functions of optimal regimes in IV settings. Here we focus on value function identification.}

Now we provide specific identification results for $g_{\mathbf{sup}}$ in the IV setting under assumptions \ref{ass: lat unconfoundedness}-\ref{ass: indep compliance type}. First, define the random variable $W(a)$ as

\begin{align*}
    W(a) \coloneqq \frac{ (2Z-1) I(A=a) Y(2a-1) }{ \delta(L) f(Z \mid L) }. 
\end{align*}

Furthermore,  let $\psi_1 (a,l)$ denote the conditional expectation of $W(a)$ given $L=l$,

\begin{align}
    \psi_1 (a,l) \coloneqq \mathbb{E} \Big[ W(a) \Big | L=l   \Big]. \label{eq: psi1}
\end{align}

Under Assumptions \ref{ass: cons}-\ref{ass: pos} and Assumptions \ref{ass: lat unconfoundedness}-\ref{ass: indep compliance type}, we have the following identification result \citep{cui2021necessary},
\begin{align}
      \mathbb{E}(Y^{a}  \mid L=l) = \psi_1 (a,l). \label{eq: psi1ID}
\end{align}

Applying \eqref{eq: psi1ID} to Lemma \ref{lemma: functional} gives a specific identification functional for $\mathbb{E}(Y^{a} \mid A = 1-a, L=l)$, as stated in the following Proposition.

    \begin{proposition}
\label{iv: identiication}
Under Assumptions \ref{ass: cons}-\ref{ass: pos} and identification Assumptions
\eqref{ass: lat unconfoundedness}-\eqref{ass: indep compliance type}, 
\begin{align*}
 \mathbb{E}(Y^{a} \mid A = a', L=l)
 & =  \begin{cases}
                    \mathbb{E}(Y \mid A = a', L=l),   & \text{if } a = a' , \\
                 \Psi(a,l)   ,  & \text{if }  a \neq a'
                \end{cases} 
\end{align*}
for all $a$, $a'$ and $l$, where 
\begin{align*}
    \Psi(a,l) &= \frac{\psi_1(a,l) - \mathbb{E}(Y | A = a, L = l)P(A = a | L = l)}{P(A= 1-a | L = l) }.
\end{align*}

Furthermore, the $L$-superoptimal regime for all $a'$ and $l$ is identified by 
\begin{align*}
     g_{\mathbf{sup}}(a',l)  
 = & \begin{cases}
                    a'   & \text{if } \ \mathbb{E}(Y \vert L = l) \geq \psi_1(1-a',l) , \\
                    1-a'  & \text{if } 
 \ \mathbb{E}(Y \vert L = l) < \psi_1(1-a',l) . 
                \end{cases}. 
\end{align*}

\end{proposition}

\subsubsection{Augmenting superoptimal regimes with instruments}
\label{app sec: IVsuperopt}

Consider an instrumental variable setting and the following regimes, which are functions of baseline covariates $L$ and the instrument $Z$.

\begin{definition}[$(L,Z)$-optimal regimes]
\label{def: z opt}

The $(L,Z)$-optimal regime, $g_{\mathbf{opt}}$, assigns treatment $A^{g_{\mathbf{opt}}+}=a$ given a vector $L=l$ and instrument $Z=z$ by
$$
   g_{\mathbf{opt}}(l,z)  \equiv \underset{a \in \{0,1\} }{\arg \max } \  \mathbb{E}(Y^a \mid L = l, Z=z). %, \text{ where $\mathcal{G}$ is the set of regimes that satisfy impartiality.}   %\nonumber \\
   $$
\end{definition}

Because $Z \independent Y^a \mid L $ under the classical IV assumptions \ref{ass: lat unconfoundedness}-\ref{ass: IV Independence}, see Appendix \ref{app sec: IVid}, the  $(L,Z)$-optimal is equal to the $L$-optimal regime; that is, 
$$
 \mathbb{E}(Y^a \mid L = l, Z=z)  =  \mathbb{E}(Y^a \mid L = l)  \ \text{ for all } a,l,z.
$$

Consider now the $(L,Z)$-superoptimal regimes.

\begin{definition}[$(L,Z)$-superoptimal regimes]
\label{def: z supopt}
The $(L,Z)$-superoptimal regime, $g_{\mathbf{sup}}$ assigns treatment $A^{g_{\mathbf{sup}}+} = a$ given $A=a'$, $L=l$ and $Z=z$ by 
 $$
  g_{\mathbf{sup}}(a',l,z)  \equiv \underset{a \in \{0,1\} }{\arg \max } \  \mathbb{E}(Y^{a} \mid A=a', L = l, Z=z). %, \text{ where $\mathcal{G}$ is the set of regimes that satisfy impartiality.}   %\nonumber \\
$$
\end{definition}
The $(L,Z)$-superoptimal regime is of particular interest when the following positivity condition is verified.
\begin{assumption}
    $P(A =a' \mid L = l,  Z = z) > 0$ for all $a',z \in \{0,1\}, l \in \mathcal{L}$.
    \label{ass: (L,Z) superopt positivity}
\end{assumption}

Furthermore, 
$$
Z \not\!\perp\!\!\!\perp Y^a \mid L,A.
$$
Indeed, there could exist $a,a',z$ and $l$ such that 
$$
 \mathbb{E}(Y^a \mid A=a', L = l, Z=z)  \neq  \mathbb{E}(Y^a \mid A=a, L = l).
$$
The implication of these results is that using the instrument as input to the decision function could further improve expected outcomes under the regime. This result is not only of theoretical interest, because under Assumptions \ref{ass: cons} and \ref{ass: lat unconfoundedness}-\ref{ass: (L,Z) superopt positivity} we can identify these regimes, as confirmed by the following proposition.

\begin{proposition}
\label{prop: functional z}
Under Assumptions \ref{ass: cons} and \ref{ass: (L,Z) superopt positivity}, $\mathbb{E}(Y^{a} \mid A = a', L=l, Z=z)$ can be expressed as 
\begin{align*}
 \mathbb{E}(Y^{a} \mid A = a', L=l, Z=z)
 & =  \begin{cases}
                    \mathbb{E}(Y \mid A = a', L=l,  Z = z),   & \text{if } a = a' , \\
                    \frac{\mathbb{E}(Y^{a} \mid L=l,  Z = z) - \mathbb{E}(Y \mid A = a, L=l,  Z = z)P(A =a \mid L = l,  Z = z)   }{P(A =a' \mid L = l,  Z = z)},  & \text{if }  a \neq a'. 
                \end{cases} 
\end{align*}
\end{proposition}
Furthermore, under Assumptions \ref{ass: lat unconfoundedness}-\ref{ass: indep compliance type}, we have that  $\mathbb{E}(Y^{a} \mid L=l,  Z = z) = \mathbb{E}(Y^{a} \mid L=l) = \psi_1(a,l)$, and  thus $\mathbb{E}(Y^a \mid A = a', L=l,  Z = z)$ is identified.

\subsection{Examples in IV settings} \label{appsec: ex}
%Short introduction to examples section
Here we give three examples where the superoptimal regime outperforms the optimal regime in IV settings. Example \ref{ex: luedkte} and Example \ref{ex: luedkte modif} are motivated by \citet{qiu2021optimal}. Example \ref{ex: iv heterogeneity} explicitly concerns a setting where the conditions from Section \ref{appsec: IV} are met.

\begin{example}
\label{ex: luedkte}
 \citet{qiu2021optimal} developed theory for identification and estimation of $L$-optimal regimes in settings with unmeasured confounding, given that a valid binary instrument is available for a binary treatment. Whereas \citet{qiu2021optimal} gave identification results for regimes that are $L$-optimal with respect to pretreatment covariates $L$, they emphasized that these $L$-optimal regimes can be worse than the regime that was implemented in the observed data \citep[Remark 1]{qiu2021optimal}, which also was illustrated by an explicit example  \citep[Remark 5]{qiu2021optimal}. Specifically:
 
 \begin{enumerate}
    \item Draw $W,U,Z \sim \text{Bernoulli} [p=0.5]$ (mutually independent draws).
    \item Draw $A^{z=0} \sim \text{Bernoulli}[p=0.3U + 0.4 ]$.
    \item Draw $A^{z=1} \sim \text{Bernoulli}[p=0.3U + 0.6 ]$.
    \item Draw $Y^{a=0} \sim \mathcal{N}[1-2U,1 ]$.
    \item Draw $Y^{a=1} \sim \mathcal{N}[2U -1,1 ]$.
\end{enumerate}
\citet{qiu2021optimal} showed that $\mathbb{E}(Y)= 0.3$ but $\mathbb{E}(Y^{g_{\mathbf{opt}}})= 0$. However, a simple computation shows that $\mathbb{E}(Y^{g_{\mathbf{sup}}})= 0.3$. Thus, the superoptimal regime is strictly better than the $L$-optimal regime. 
 
\end{example}

\begin{example}
\label{ex: luedkte modif}
We also consider a modified version of Example \ref{ex: luedkte}, where only $P(A^{z} =a \mid L,U,Z )$ is changed, that is:
\begin{enumerate}
    \item Draw $W,U,Z$ $ \sim \text{Bernoulli} [p=0.5]$ (mutually independent draws).
    \item Draw $A^{z=0} \sim 1 - \text{Bernoulli}[p=0.3U + 0.4 ]$.
    \item Draw $A^{z=1} \sim 1 - \text{Bernoulli}[p=0.3U + 0.6 ]$.
    \item Draw $Y^{a=0} \sim \mathcal{N}[1-2U,1 ]$.
    \item Draw $Y^{a=1} \sim \mathcal{N}[2U -1,1 ]$.
\end{enumerate}

%Because everything except $P(A^{z}=a \mid W,U,Z )$ is identical to Example \ref{ex: luedkte}, 
The $L$-optimal regime  in this example is equal to the $L$-optimal regime in Example \ref{ex: luedkte}; that is, $\mathbb{E}(Y^{g_{\mathbf{opt}}})= 0$; only steps 2. and 3. are changed from Example \ref{ex: luedkte}. However,  a simple calculation shows that $\mathbb{E}(Y)= - 0.3$, and, thus, the observed regime is inferior to the $L$-optimal regime. Another simple calculation shows that $\mathbb{E}(Y^{g_{\mathbf{sup}}})= 0.3$, which implies that the $L$-superoptimal is strictly better than the observed regime and the $L$-optimal regime. Thus, this example illustrates that the $L$-superoptimal regimes can be (strictly) better than the optimal regime or regimes that improve on a baseline regime, such as the no-treatment ($a = 0$) or observed regimes \citep{kallus2021minimax,cui2021semiparametric,qiu2021optimal}.
\end{example}

\begin{example}
\label{ex: iv heterogeneity}
 IV assumptions \ref{ass: lat unconfoundedness}-\ref{ass: indep compliance type} used to identify $L$-superoptimal regimes do impose restrictions on the counterfactual outcome distributions. However, in the following simple example, we nevertheless illustrate that these restrictions do not preclude the $L$-superoptimal regime from being better than the $L$-optimal regime. 

Suppose IV assumptions \ref{ass: lat unconfoundedness}-\ref{ass: IV Independence} hold. Let $L = \emptyset$, and consider $U,Z,A,Y \in \{0,1\}$ such that 
\begin{itemize}
    \item $P(U=1 )=P(Z=1) = 0.5$
    \item $P(A=1 \mid Z=z, U = u) = 0.5c + cz + cu, $ where $0 < c < 0.4$
    \item $\mathbb{E}(Y^{a=1} ) = 0.5 $, $\mathbb{E}(Y^{a=1} \mid U =1 ) = 0.1 $, $\mathbb{E}(Y^{a=1} \mid U =0 ) = 0.9 $ 
    \item $\mathbb{E}(Y^{a=0} ) = 0.4 $, $\mathbb{E}(Y^{a=0} \mid U =1 ) = 0.7 $, $\mathbb{E}(Y^{a=0} \mid U =0 ) = 0.1 $ 
\end{itemize}
This example is consistent with the SWIG in Figure \ref{fig:superoptimal} if we had removed $L$ from the graph. Trivially, $g_{\textbf{opt}}=1$. To calculate $g_{\textbf{sup}}$, we use law of total probability to find the values of $P(U=u \mid Z=z, A = a')$ for all $u,z$ and $a'$, and then compute $\mathbb{E}(Y^{a} \mid A=a', Z = z ) $ from 
$$
\mathbb{E}(Y^{a} \mid A=a', Z = z ) = \sum_u \mathbb{E}(Y^{a} \mid U=u ) P(U=u \mid A=a', Z = z ).
$$
In particular, for $c=\frac{1}{5}$ we find that 
\begin{itemize}
    \item $\mathbb{E}(Y^{a=1} \mid A=1, Z = 1 ) = {4}/{10} < \mathbb{E}(Y^{a=0} \mid A=1, Z = 1 ) = {19}/{40}$, 
    \item $\mathbb{E}(Y^{a=1} \mid A=0, Z = 0 ) = {11}/{20} > \mathbb{E}(Y^{a=0} \mid A=0, Z = 0 ) = {17}/{80}$.
\end{itemize}
 Thus, $g_{\textbf{sup}} (a'=1,z=1) = 0 \neq g_{\textbf{sup}} (a'=0,z=0) = 1$, which is sufficient to illustrate that the superoptimal regime is better than the optimal regime. 
\end{example}

\subsection{A test for unmeasured confounding} \label{appsec: test}
Here we outline a strategy to construct hypothesis tests that can detect unmeasured confounding. The following proposition follows by the same logic of Corollary \ref{cor: equiv} in the main text, and states constraints on counterfactual parameters that can be used to create such tests. 

\begin{proposition} \label{prop: UC}

Let $\Pi \equiv \{g: \mathcal{L} \rightarrow \{0,1\}\}$ be the class of regimes $g$ that depend only on $L$, and let $\Pi^{*}$ be the analogous class of regimes $g^*$ that may depend additionally on $A$. Then consider $B$ and $C$ to be coarsenings of $A$ of $L$, respectively. Define the interval $\mathcal{I}^g_c$ constructed from the following upper and lower bounds:
\begin{align*}
    \underset{g\in\Pi}{\max } & \ \mathbb{E}(Y^g \mid C= c) \text{ and } \\
    \underset{g\in\Pi}{\min } & \ \mathbb{E}(Y^g \mid  C= ).
\end{align*}
Likewise define the interval $\mathcal{I}^{g^*}_{b,c}$ to be the analogous interval constructed by minimizing and maximizing over $g^*\in\Pi^*$, and additionally conditioning on $B=b$. 

Suppose $Y^a \independent A \mid L$ for $a\in\{0,1\}$. Then, for each $c$, 
$$
    \mathcal{I}^{g^*}_{b,c} \equiv  \mathcal{I}^{g}_{c} \text{ for all } b.
$$
\end{proposition}

\begin{corollary}
\label{cor: test}
If  $Y^a \independent A \mid L$ then $\mathbb{E}(Y \mid  B=b, C=c)$ and $\mathbb{E}(Y^{g_{\mathbf{sup}}} \mid  B=b, C=c)$ are contained in the interval $\mathcal{I}^g_{c}$ for all $b,c$.
\end{corollary}
Corollary \ref{cor: test} follows immediately because $Y^{g_{\mathbf{sup}}}$ is obviously an element in $\Pi^*$ and also because $Y=Y^{g^*}$ w.p.1 for the special regime in $\Pi^*$ that assigns $A^{g+}=A$. 

When the conditions of Proposition \ref{thm: id} hold, then $\mathcal{I}^g_{c}$ is identified, as are the conditional average outcomes under the superoptimal regime $g_{\mathbf{sup}}$. Thus, we can re-express Proposition \ref{prop: UC} in terms of constraints on the law of the observed data. Consideration of the contrapositive of this re-expressed Proposition \ref{prop: UC} then ostensibly motivates a simple test for unmeasured confounding: if any of the constraints are violated, then $Y^a \not\independent A \mid L$ for some  $a \in \{0,1\}.$ The practical utility of this test, however, is questionable, because it will have low (or zero) power in many settings. In particular, common identification strategies, such as those based on the g-formula \citep{robins1986new}, would rely on no unmeasured confounding assumptions for identification, e.g.\ $Y^a \independent A \mid L, \ a \in \{0,1\}.$ Whenever such conditions are (possibly erroneously) assumed, the test has zero power. In IV settings, however, the test could reject the null hypothesis of no unmeasured confounding. We illustrate one such setting in Example \ref{ex: luedkte}. Thus, this simple test adds to the small number of existing tests for unmeasured confounding, which, to our knowledge, critically rely on instrumental variables \citep{guo2014using,de2014testing}.

%let $h(c)$  be an identification functional for $ {\mathbb{E}}(Y^{g*} \mid C = c)$, where $c \in \mathcal{C} \subseteq \mathcal{L}$. If there exists $c \in \mathcal{C}$ such that $\mathbb{E}(Y \mid C = c ) \geq h(c)$, then there is unmeasured confounding between $A$ and $Y$. 

\subsection{Estimation under an IV model} \label{appsec: est}
%Short introduction to testing section - give disclaimers
We can leverage classical results on semi-parametric estimation to construct estimators of outcomes under superoptimal regimes in IV settings. In particular, we build on existing results by \citet{cui2021semiparametric} to construct a multiply robust estimator. We also construct a simple semi-parametric estimator, which we implemented in the application of Section \ref{sec: icu example}. 
\subsubsection{An influence function for the $(L,A)$-CATE}
\label{appsec: influence function id}

Consider the observed data $\mathcal{O} = (Y,A,L,Z)$ in our instrumental variable setting of Appendix \ref{app sec: IVid}, and suppose that $\mathcal{O}$ is described by the law $P$ that belongs to the model $\mathcal{M}=\{P_\theta:\theta\in \Theta\}$, where $\Theta$ is the parameter space. We will study the efficient influence function $\varphi^{\eff}(\mathcal{O})$ for a parameter $\Psi\equiv \Psi(\theta)$ in a non-parametric model $\mathcal{M}_{\text{np}}$ that imposes no restrictions on $P$ except positivity, such that $\varphi^{\eff}(\mathcal{O})$ is given by
${d\Psi(\theta_t)}/{dt}\vert_{t=0} = E\{\varphi^{\eff}(\mathcal{O})S(\mathcal{O})\}$, where ${d\Psi(\theta_t)}/{dt}\mid_{t=0}$ is the pathwise derivative of $\Psi$ along any parametric submodel of $P$ indexed by $t$, and $S(\mathcal{O})$ is the score function of the parametric submodel evaluated at $t=0$, see more details in, for example,  \citet{newey1994} and \citet{Van2000}. First, we will build on results from \citet{cui2021semiparametric} who  derived the efficient influence function for $ \psi_1(a,l)$, which we denote by $ \psi_1^{ \text{eff}}(a,l)$ in the following proposition. 

\begin{proposition}
\label{thm:proof eff infl}
Suppose covariates $L$ are discrete. The efficient influence function of $\Psi(a,l)$  in $\mathcal{M}_{\text{np}}$ is given by 
\begin{align*}
    &  \Psi^{\text{eff}}(a,l) \color{black} = \\
    & \frac{1}{P( A = 1-a \mid  L = l)^2}\Bigg(\psi_1^{\text{eff}}(a,l)P( A = 1-a  \mid L = l)\\
    &- \psi_1(a,l) \frac{I[L = l]}{P(L = l)} (I[A = 1-a] - P(A = 1-a  \mid  L = l))\\
    & - \bigg(\frac{I[A = a, L = l]}{P(A = a ,L = l)}(Y - \mathbb{E}(Y  \mid  A = a, L = l)) P(A = a  \mid  L = l)  \\
    & + \mathbb{E}(Y  \mid  A = a, L = l)\frac{I[L = l]}{P(L = l)}(I[ A= a] - P( A = a  \mid  L = l))\bigg)P(A = 1-a  \mid  L = l) \\
    & + \mathbb{E}(Y  \mid  A = a, L = l) P( A = a  \mid  L = l) \frac{I[L = l]}{P(L = l)} (I[A = 1-a] - P(A = 1-a  \mid  L = l)) \Bigg),
\end{align*}
where 
\begin{align*}
 \psi_1^{\text{eff}}(a,l) =  &\frac{(2Z-1) (2A - 1) Y I[A = a]}{\delta(l)f(Z \mid l)P(L = l)} - \bigg(\frac{Z \mathbb{E}((2A - 1) Y I[A = a] \mid Z,L= l)}{\delta(l)f(Z \mid l) P(L = l)} \nonumber \\
    & - \sum_z \frac{z \mathbb{E}((2A-1)YI[A = a] \mid Z = z, L = l)}{\delta(l)P(L = l)} \nonumber \\
    & + \frac{(2Z-1)[(2A-1) - \mathbb{E}(2A - 1 \mid Z,L=l)]}{2 f(Z \mid l) \delta(l)P(L = l)}\\
    &\cdot \sum_z\frac{z \mathbb{E}((2A-1)YI[A = a] \mid Z = z, L = l)}{\delta(l)} \bigg) \nonumber \\
    & - \mathbb{E}\left[\frac{(2Z- 1) (2A - 1) Y I[A = a]}{\delta(l)f(Z \mid l)P(L = l)} \mid L = l  \right] ,
\end{align*}
that is, $ \psi_1^{\text{eff}}(a,l)$ is the nonparametric influence function for $\psi_1(a,l)$.
\end{proposition}

Let $\mathbb{P}_n$ be the empirical distribution of the observed data $\mathcal{O}$; we write $\mathbb{P}_n(T(\mathcal{O}))$ to be the empirical mean of $T(\mathcal{O})$. The influence function in Proposition \ref{thm:proof eff infl} motivates a one-step estimator of $\Psi(a,l)$,  that is,  
\begin{align*}
    & \frac{\hat{\psi}_1(a,l) - \hat{\mathbb{E}}(Y  \mid A = a, L = l) \hat{P}(A = a  \mid  L = l)}{\hat{P}(A = 1-a  \mid  L = l)}\\
    & + \mathbb{P}_n \Bigg(\frac{1}{\hat{P}( A = 1-a \mid  L = l)^2}\Bigg(\hat{\psi}_1^{\text{eff}}(a,l) \hat{P}( A = 1-a  \mid L = l) - \hat{\psi}_1(a,l) \frac{I[L = l]}{\hat{P}(L = l)} (I[A = 1-a] \\
    & - \hat{P}(A = 1-a  \mid  L = l))- \bigg(\frac{I[A = a, L = l]}{\hat{P}(A = a ,L = l)}(Y - \hat{\mathbb{E}}(Y  \mid  A = a, L = l)) \hat{P}(A = a  \mid  L = l)  \\
    & + \hat{\mathbb{E}}(Y  \mid  A = a, L = l)\frac{I[L = l]}{\hat{P}(L = l)}(I[ A= a] - \hat{P}( A = a  \mid  L = l))\bigg)\hat{P}(A = 1-a  \mid  L = l) \\
    & + \hat{\mathbb{E}}(Y  \mid  A = a, L = l) \hat{P}( A = a  \mid  L = l) \frac{I[L = l]}{\hat{P}(L = l)} (I[A = 1-a] - \hat{P}(A = 1-a  \mid  L = l)) \Bigg) \Bigg).
\end{align*}
This estimator requires the following models to be correctly specified:
\begin{itemize}
    \item the model for $ \psi_1(a,l)  $,
    \item the model for $P(A = a^\dagger  \mid  L = l)$, and
    \item the model for $\mathbb{E}(Y  \mid  A = a , L = l)$.
\end{itemize}

Semi-parametric multiple robust estimators for expressions similar to $\psi_1(a,l)$ have been suggested, see e.g.\ \citet{cui2021semiparametric}. These estimators can be used when estimating $\Psi(a,l)$. In Appendix \ref{appsec: robust semiparam estim for CATE} we also give the efficient influence function for an expression that identifies $L$-superoptimal regimes, without identifying the value function. 

%Furthermore, the results in this section motivate a simple semi-parametric estimator that can be used to estimate value functions under both $L$-optimal and $L$-superoptimal regimes with standard regression models when $\Psi(a,l)$ identifies $\mathbb{E}(Y^{a}  \mid A=1-a, L)$, as stated in Appendix \ref{appsec: estim algorithm}.

In this subsection, we derive the influence function of the marginal value function $\mathbb{E}(Y^{g(a,l)})$. Let $g(a,l)$ be a known regime that is a function of $a$ and $l$. 
First,
\begin{align*}
    \mathbb{E}(Y^{g(A,L)}) &= \sum_{a,l} \mathbb{E}(Y^{g(a,l)}  \mid  A = a, L = l) P(A = a, L = l).
\end{align*}
Let $\mathbb{IF}(\mathbb{E}(Y^{g(a,l)}))$ denote the influence function of $\mathbb{E}(Y^{g(a,l)})$. As $g(a,l)$ is constant given $a$ and $l$, we can use differentiation rules to obtain
\begin{align*}
    \mathbb{IF}(\mathbb{E}(Y^{g(a,l)})) = \sum_{a,l}  &I[g(a,l) = a] \frac{I[A = a, L = l]}{P(A = a, L = l)}(Y - \mathbb{E}(Y  \mid  A = a, L = l))P(A = a, L = l)\\
    & + I[g(a,l) = 1 - a] \Psi^{\text{eff}}(1-a,l)P(A = a,L =l)\\
    & + \mathbb{E}(Y^{g(a,l)}  \mid  A = a, L = l)(I[A = a, L =l] - P(A = a, L = l)) ,
\end{align*}
assuming $L$ is discrete. The argument extends to continuous (real-valued) $L$ by replacing sums with integrals. 

\subsubsection{A robust semi-parametric efficient estimator of the $(L,A)$-CATE}
\label{appsec: robust semiparam estim for CATE}

Estimating the $L$-superoptimal regime $g_{\mathbf{sup}}$ often requires fewer model assumptions than estimating the value function $\mathbb{E}(Y^{g_{\mathbf{sup}}})$. Indeed, it follows from Corollary \ref{cor: sign id} that we only need to identify and estimate the (sign of) the difference $\mathbb{E}(Y \mid L=l)-\psi_1$, which has the non-parametric influence function stated in the following corollary: \\

\begin{corollary}
\label{cor: sign id}
The efficient influence function of $\mathbb{E}(Y \mid L=l)-\psi_1$  in $\mathcal{M}_{\text{np}}$ is given by

\begin{align*}
    \frac{I( L = l)}{P(L = l)}(Y - \mathbb{E}(Y \mid L = l)) - \psi_1^{\text{eff}}.
\end{align*}
\end{corollary}
Corollary \ref{cor: sign id} follows immediately from Proposition \ref{thm:proof eff infl}. As $L$ is discrete, the influence function of $\mathbb{E}(Y  \mid  L =l)$ is
\begin{align*}
    \frac{I( L = l)}{P(L = l)}(Y - \mathbb{E}(Y \mid L = l)).
\end{align*}
Using differentiation rules, we find that the influence function of $\mathbb{E}(Y  \mid  L = l) - \mathbb{E}(Y^a  \mid  L = l)$ is
\begin{align*}
    \frac{I( L = l)}{P(L = l)}(Y - \mathbb{E}(Y \mid L = l)) - \psi_1^{\text{eff}}.
\end{align*}

A one-step estimator of $\mathbb{E}(Y \mid L = l) - \mathbb{E}(Y^a  \mid  L = l)$ is \\
\begin{equation}
    \hat{\mathbb{E}}(Y  \mid  L = l) - \hat{\psi}_1 + \mathbb{P}_n\left(\frac{I( L = l)}{\hat{P}(L = l)}(Y - \hat{\mathbb{E}}(Y \mid L = l)) - \hat{\psi}_1^{\text{eff}} \right). \label{eq: one-step estimator app C}
\end{equation}

The one-step estimator in Equation \eqref{eq: one-step estimator app C} is the sum of a one-step estimator of the conditional mean $\mathbb{E}(Y| L = l)$ and the one-step estimator of $\psi_1$ derived by \citet{cui2021semiparametric}. This one-step estimator has robustness properties; if there exists $\epsilon > 0$ such that $\hat{P}(L = l) > \epsilon$ w.p.1, and one of the following combinations of models required for consistency of the one-step estimator of $\mathbb{E}[Y | L =l]$ and the one-step estimator of $\psi_1$ \citep{cui2021semiparametric} are correctly specified, 
\begin{itemize}
    \item $\mathcal{M}_1$: models for $f(Z  \mid  L)$, $\delta(L)$, and $P(L = l)$,
    \item $\mathcal{M}_2$: models for $f(Z  \mid  L)$,  \\  $\gamma(L) \coloneqq \sum_z \{ (z + 1)/2 \mathbb{E}((A+1)Y I\{A = a\}/2  \mid  Z = z, L = l) \}/ \delta(L)$, and $P(L = l)$
    \item $\mathcal{M}_3$: models for $\gamma(L)$, $\gamma'(L) \coloneqq \mathbb{E}((A + 1)Y I\{A = a \}/2  \mid  Z = 0, L )$, $\delta(L)$, $\mathbb{E}(A  \mid  Z = 0,L)$, and $P(L = l)$
    \item $\mathcal{M}_4$: models for $f(Z  \mid  L)$, $\delta(L)$, and $\mathbb{E}[Y| L = l]$,
    \item $\mathcal{M}_5$: models for $f(Z  \mid  L)$,  \\  $\gamma(L) \coloneqq \sum_z \{ (z + 1)/2 \mathbb{E}((A+1)Y I\{A = a\}/2  \mid  Z = z, L = l) \}/ \delta(L)$, and $\mathbb{E}[Y | L = l]$
    \item $\mathcal{M}_6$: models for $\gamma(L)$, $\gamma'(L) \coloneqq \mathbb{E}((A + 1)Y I\{A = a \}/2  \mid  Z = 0, L )$, $\delta(L)$, $\mathbb{E}(A  \mid  Z = 0,L)$, and $\mathbb{E}[Y| L = l]$
\end{itemize}
the one-step estimator of Equation \eqref{eq: one-step estimator app C} is correctly specified.

Under standard regularity conditions, the one-step estimator is consistent and asymptotically normal under the union model of $\mathcal{M}_1$ to $\mathcal{M}_6$ \citep{cui2021semiparametric}.

\subsubsection{Implementing a simple semi-parametric estimator of the $(L,A)$-CATE}
\label{appsec: estim algorithm}

Here we describe the semi-parametric estimator that is used in our application in Section \ref{sec: icu example}. Suppose that Assumptions \ref{ass: cons}-\ref{ass: pos}, and Assumptions \ref{ass: lat unconfoundedness}-\ref{ass: indep compliance type} of Appendix \ref{app sec: IVid} hold. Under these assumptions, the identification result in Proposition \ref{iv: identiication} is valid. We use Greek letters to denote parameters indexing statistical models, and we use hats to denote estimates. 
As before, let $W(a) = \frac{ (2Z-1) Y(2A-1) I(A=a)  }{ \delta(L) f(Z \mid L) }, $ and $ \psi_1(a,l) = \mathbb{E}(W(a) \mid L=l) $. Then the following estimation algorithm for the value function of estimates of the optimal or superoptimal regimes is consistent when the parametric regression models are correctly specified (see Appendix \ref{appsec: estimproofs} for a proof):
\begin{enumerate}
  %  \item Using data from all individuals, compute an estimate $f(Z \mid L; \alpha)$ of $f(Z \mid L)$ by fitting a (parametric) regression model, e.g.\ logistic regression with dependent variable $Z$ and independent variables a specified function of $L$. 
    \item Using data from all individuals, compute an estimate of $\delta(L)$ by
    \begin{itemize}
        \item first computing the MLE $\hat \beta_z$ for $z=0,1$ using the (parametric) models $P(A=1  \mid  Z=1,L; \beta_1)$ and $P(A=1  \mid  Z=0,L; \beta_0)$ of $P(A=1  \mid  Z=z,L)$, e.g.\ a logistic regression model with the dependent variable $A$ and the independent variable a specified function of $Z$ and $L$.
        \item Compute an estimate $\delta(L; \hat \beta_0,\beta_1) = P(A=1  \mid  Z=1,L; \hat \beta_1)-P(A=1  \mid  Z=0,L; \hat \beta_0) $.
    \end{itemize}
    \item Using data from all individuals, compute an estimate $\hat{\mathbb{E}}\big( W(a) \mid L   \big)$ of $\mathbb{E}\big( W(a) \mid L   \big)$ as follows:
    \begin{enumerate}
        \item For each $a \in \{0,1\}$, compute the MLE $\hat \theta_{az}$ of $\theta_{az}$ for $a,z \in \{0,1\}$ for the model $\mathbb{E}\big( Y(2A-1)I(A=a) \mid L=l,Z=z ; \theta_{az} \big)$ of $\mathbb{E}\big( Y(2A-1)I(A=a) \mid L=l,Z=z \big)$, which we will use to estimate $\psi_1(a,l) = \mathbb{E}(W(a) | L = l)$, with independent variables $L$; that is, fit models
        \begin{align*}
            & \mathbb{E}\big( -YI (A=0) \mid L=l,Z=0 ; \theta_{00} \big), \\
            & \mathbb{E}\big( -YI (A=0) \mid L=l,Z=1 ; \theta_{01} \big) , \\
            & \mathbb{E}\big( YI(A=1) \mid L=l,Z=0; \theta_{10} \big) , \\
            & \mathbb{E}\big( YI(A=1) \mid L=l,Z=1; \theta_{11} \big) . 
        \end{align*}
        When $A$ and $Y$ are binary, then $\mathbb{E}\big( -YI (A=0) \mid L=l,Z=z \big) + 1 \ \in \ \{0,1\}$ and $\mathbb{E}\big( YI(A=1) \mid L=l,Z=z \big) \ \in \ \{0,1\}$. We can thus fit regression models for binary outcomes to estimate each of the four conditional expectations, and then we compute estimates $\hat{\psi}_1(a,l) \coloneqq  \hat{\mathbb{E}}\big[ W(a) \mid L=l   \big]$ of ${\psi}_1(a,l)$ by
        $$
       \hat{\psi}_1(a,l) =  \sum_z \frac{ (2z-1)  }{ \hat \delta(l) } \mathbb{E}\big( Y(2A-1)I(A=a) \mid L=l,Z=z; \hat \theta_{az} \big),
        $$
        where $\hat \theta$ is the MLE.%\footnote{When $Y$ is binary we know that, for each $a$, $ 0 \leq \mathbb{E}({W}(a) \mid L) \leq 1 $ w.p.1.}  
        % \item     a (parametric) regression model, e.g.\  with dependent variable $$\hat{W}^{(1)} = \frac{ (2Z-1) Y(2A-1) }{ {\delta}(L; \beta) f(Z \mid L; \alpha) } ,$$ and independent variable  a specified function of $L$. However, when $Y$ i binary we know that $ 0 \leq \mathbb{E}({W}^{(1)} \mid L) \leq 1 $ w.p.1. Then, we back transform  \citet{wang2018bounded}[Equation (10)] suggest a bounded IPW estimator which seems to work well in 
    \end{enumerate}

    \item Using data from all individuals, compute the MLE $\hat \gamma$ of $\gamma$ for a model $\mathbb{E}(Y \mid L=l; \gamma) $ of $\mathbb{E}(Y \mid  L=l) $. If $Y$ is binary, use e.g.\ a logistic regression with dependent variable $Y$ and independent variables a specified function of $L$.  \item For all $a',l$, the estimated superoptimal regime $\hat{g}_{\mathbf{sup}}(a',l)$ is
    $$
\hat{g}_{\mathbf{sup}}(a',l) = \underset{a }{\arg \max }  \  I[a = a'] \mathbb{E}(Y \mid  L=l; \hat \gamma) + I[a \neq a'] \hat{\psi}_1(a,l) , \
$$
where $\mathbb{E}(Y \mid  L=l; \hat \gamma)$ and $\hat{\psi}_1(a,l) = \hat{\mathbb{E}}\big[ W(a) \mid L=l  \big]$ are estimated expected conditional means from the models specified earlier.
\begin{itemize}
    \item Similarly, if the interest is the regime ${g}_{\mathbf{opt}}(l)$,
    $$
     \hat{g}_{\mathbf{opt}}(a',l)  = \hat{g}_{\mathbf{opt}}(l) = \underset{a}{\arg \max }  \   \hat{\psi}_1(a,l) , \
    $$
    because $\hat{\psi}_1(a,l)$ can be interpreted as an estimator of $\mathbb{E}(Y^a \mid L=l)$, as showed in \cite{cui2021semiparametric}.
\end{itemize}
 \item 
We estimate ${\mathbb{E}}(Y^{\hat{g}_{\mathbf{sup}}} )$ by
\begin{align*}
    & \hat {\mathbb{E}}(Y^{\hat{g}_{\mathbf{sup}}} ) = \\
&  \mathbb{P}_n \Bigg\{ I[A \neq  \hat{g}_{\mathbf{sup}}(A,L) ] \hat{V}(A, L) + I[A =  \hat{g}_{\mathbf{sup}}(A,L)]\hat{\mathbb{E}}(Y \mid A, L) ]  \Bigg\},
\end{align*}
where
\begin{align}
\hat{V}(a,l) &:= \frac{\hat{\psi}_1(1-a,l) - \hat{\mathbb{E}}(Y | A = 1-a,L =l)\hat{P}(A = 1-a | L = l)}{\hat{P}(A = a| L = l)},
\end{align}
and $\hat{\mathbb{E}}(Y | A = 1 - a,L)$ are $\hat{P}(A = 1| L = l)$ are estimated with some chosen models.

Alternatively, we could estimate ${\mathbb{E}}(Y^{\hat{g}_{\mathbf{sup}}} )$ by 
\begin{align*}
& \hat {\mathbb{E}}(Y^{\hat{g}_{\mathbf{sup}}} ) = \\
&  \mathbb{P}_n \Bigg\{ I[A \neq  \hat{g}_{\mathbf{sup}}(A,L) ]  \bigg\{ \frac{\hat{\mathbb{E}}\big( W(A \mid L  \big) - \hat{\mathbb{E}}(Y|A,L)\hat{\pi}(A|L)}{\hat{\pi}(A|L)}  \bigg\}
+ I[A =  \hat{g}_{\mathbf{sup}}(A,L) ] Y \Bigg\},
\end{align*}
where $\hat{\pi}(a | l)$ is an estimate of $P(A = a | L = l)$.

\item To obtain confidence intervals for the value function,
\begin{enumerate}
    \item estimate $\hat{g}_{\mathbf{sup}}(a,l) $ from the entire data, and
    \item for each bootstrap sample $b$, estimate $\hat {\mathbb{E}}_b(Y^{\hat{g}_{\mathbf{sup}}}  )$ as in step 5, where subscript $b$ indicates the bootstrap sample. 
\end{enumerate}
\end{enumerate}
We show that this algorithm is a consistent and asymptotically normal estimator of the value function of a fixed regime $g$, in this case the estimated optimal or superoptimal regimes, under suitable conditions in Appendix \ref{appsec: estimproofs}. Step 5 of the estimation algorithm is derived from an identification result that we explicilty state in Lemma \ref{lemma: Id formula of E(Y^g)} of Appendix \ref{appsec: estimproofs}. Furthermore, Appendix \ref{cor: bootstrap justification} justifies the use of the bootstrap procedure in Step 6.

In our data example in Section \ref{sec: icu example} of the main text we performed the algorithm on sample-split data; that is, we estimated the optimal regime in 60\%  of the data, and subsequently estimated the value function under this \textit{estimated} optimal regime with confidence intervals in the remaining 40\% of the data. We parameterized the following terms with logistic regression models:
\begin{itemize}
    \item $P(A = a | L = l)$,
    \item $P(A = 1 | Z = z, L = l)$,
    \item $\mathbb{E}(-YI[A = 0] | L = l, Z = 0)$,
    \item $\mathbb{E}(-YI[A = 0] | L = l, Z = 1)$,
    \item $\mathbb{E}(YI[A = 1] | L = l, Z = 0)$,
    \item $\mathbb{E}(YI[A = 1] | L = l, Z = 1)$,
    \item $\mathbb{E}(Y | A = a, L = l)$, and
    \item $\mathbb{E}(Y | L =l)$.
\end{itemize}
We used percentile bootstrap confidence intervals and truncated them to the boundary of the parameter space. 

Furthermore, we repeated our analysis with different sample splits, and these analyses gave similar results with overlapping confidence intervals. However, when using some splits, for example when we estimated the optimal regime in 30\%  of the data and subsequently estimated the value function with confidence intervals in the remaining 70\% of the data, we found that the estimated regimes were affected by sampling variability and overfitting, as discussed in Section \ref{sec: algorithm ass decision making}. In particular, the estimated $L$-superoptimal regimes ($\hat g_{\mathbf{sup}}$)  gave point estimates larger than the estimated $(L,Z)$-superoptimal regimes ($\hat g_{\mathbf{z-sup}}$), although the confidence intervals were overlapping.

Here we show that the semi-parametric  algorithm is an asymptotically normal estimator of $\mathbb{E}(Y^{g})$ for a given regime $g$ under suitable conditions. This algorithm is not restricted to the models we use in Appendix \ref{appsec: estim algorithm} and in the example of Section \ref{sec: icu example}. Hence, we denote the estimated means and probabilities with hats, that is, we write $\hat{\mathbb{E}}$ and $\hat{P}$, and specify the conditions required for convergence in Proposition \ref{Prop: Asympt Normal} and Remark \ref{rem: relaxation of Donsker} of this section.

\begin{proposition}
Under Assumptions \ref{ass: cons}-\ref{ass: pos}, if
\begin{enumerate}
\item $Y$ is bounded, \label{Ass: Bounded Y}
\item $\delta(L)$ is bounded away from $0$ w.p.1,
\item $\hat{P}(A = a | L = l) \in (\epsilon_1, 1-\epsilon_1)$ w.p.1 for some $\epsilon_1 > 0$,
\item $\hat{\delta(L)} \in (\epsilon_2, 1-\epsilon_2)$ w.p.1 for some $\epsilon_2 > 0$,
\item $\hat{\mathbb{E}}(Y | A,L), \hat{P}(A = a | Z, L), \hat{\mathbb{E}}(Y(2A-1)I(A = a) | L,Z)$, and $\hat{P}(A = a | L )$ are $P$-Donsker with vanishing variance, then \label{Ass: Donsker prop}
\end{enumerate}

\begin{align*}
\sqrt{n}(\hat{\mathbb{E}}(Y^g) - \mathbb{E}(Y^g)) \to^d N(0,\sigma^2),
\end{align*}
for some $\sigma^2 < \infty$ as $n \to \infty$.

\label{Prop: Asympt Normal}
\end{proposition}

Conditions \ref{Ass: Bounded Y}-\ref{Ass: Donsker prop} of Proposition \ref{Prop: Asympt Normal} are standard in the optimal regime estimation literature, see for example \citet{luedtke2016optimal}. Condition \ref{Ass: Donsker prop} can be relaxed to include $\sqrt{n}$-consistent estimators if we estimate the parameters in a different sample, see for example \citet{Chernozhukov2018}.
\label{rem: relaxation of Donsker}

Here we argue why we can use the non-parametric bootstrap to compute confidence intervals for the estimator described in Section \ref{appsec: estim algorithm}, when the models are correctly specified. By the proof of Proposition \ref{Prop: Asympt Normal}, 
\begin{align*}
&\hat{\mathbb{E}}(Y^g)\\
&=  \mathbb{P}_n (I[g(A,L) \neq A] V(A,L) + I[g(A,L) = A]\mathbb{E}(Y |A, L))\\
& - \mathbb{E}(I[g(A,L) \neq A)V(A,L) + I[g(A,L) = A]\mathbb{E}(Y |A, L)]  + \mathbb{E}(Y^g) + o_P(n^{-1/2}),
\end{align*}
is the sum of a sample mean, some constants, and a vanishing term of order $\sqrt{n}$. Therefore, the bootstrap procedure can be used to derive confidence intervals \citep[Chapter 2]{Davison_Hinkley_1997}.
\label{cor: bootstrap justification}

\subsection{Proofs} \label{appsec: proofs}
Here we give proofs of Propositions \ref{iv: identiication}, \ref{prop: UC}, \ref{thm:proof eff infl} and \ref{Prop: Asympt Normal}. Some of these proofs follow from existing results, which are cited. %Our proof of the influence function based estimator builds on classical semi-parametric theory and, in particular, the estimator suggested by \citep{cui2021semiparametric}. 
\subsubsection{Identification proofs} \label{appsec: IVidproofs}
%IV id proof
{
\renewcommand{\theproposition}{}
\begin{proposition*}[Re-statement of Proposition \ref{iv: identiication}]
Under Assumptions \ref{ass: cons}-\ref{ass: pos} and identification Assumptions
\eqref{ass: lat unconfoundedness}-\eqref{ass: indep compliance type}, 
\begin{align*}
 \mathbb{E}(Y^{a} \mid A = a', L=l)
 & =  \begin{cases}
                    \mathbb{E}(Y \mid A = a', L=l),   & \text{if } a = a' , \\
                 \Psi(a,l)   ,  & \text{if }  a \neq a'
                \end{cases} 
\end{align*}
for all $a$, $a'$ and $l$, where 
\begin{align*}
    \Psi(a,l) &= \frac{\psi_1(a,l) - \mathbb{E}(Y | A = a, L = l)P(A = a | L = l)}{P(A= 1-a | L = l) }.
\end{align*}

Furthermore, the $L$-superoptimal regime for all $a'$ and $l$ is identified by 
\begin{align*}
     g_{\mathbf{sup}}(a',l)  
 = & \begin{cases}
                    a'   & \text{if } \ \mathbb{E}(Y \vert L = l) \geq \psi_1(1-a',l) , \\
                    1-a'  & \text{if } 
 \ \mathbb{E}(Y \vert L = l) < \psi_1(1-a',l) . 
                \end{cases}. 
\end{align*}

\end{proposition*} 
%The following proof relies on the identification result in Equation \eqref{eq: psi1ID}, which is derived by \citet{cui2021semiparametric}.
\begin{proof}
    %The first statement about $\mathbb{E}[Y^a | A = a', L = l]$ follows from Lemma \ref{lemma: functional} and Assumptions \ref{ass: lat unconfoundedness}-\ref{ass: indep compliance type}, which imply $\mathbb{E}[Y^{a}| L = l] = \psi_1(a,l)$ \citep{cui2021necessary}.
Plugging in Equation \eqref{eq: psi1ID} into Lemma \ref{lemma: functional}, and using Definition \ref{def: supopt} and simple algebra, we obtain the result; see Corollary \ref{cor: sign superopt} for an equivalent statement.

\end{proof}
}

%Testing proof

{
\renewcommand{\theproposition}{}
\begin{proposition*}[Re-statment of Proposition \ref{prop: UC}]

Let $\Pi \equiv \{g: \mathcal{L} \rightarrow \{0,1\}\}$ be the class of regimes $g$ that depend only on $L$, and let $\Pi^{*}$ be the analogous class of regimes $g^*$ that may depend additionally on $A$. Then consider $B$ and $C$ to be coarsenings of $A$ of $L$, respectively. Define the interval $\mathcal{I}^g_c$ constructed from the following upper and lower bounds:
\begin{align*}
    \underset{g\in\Pi}{\max } & \ \mathbb{E}(Y^g \mid C= c) \text{ and } \\
    \underset{g\in\Pi}{\min } & \ \mathbb{E}(Y^g \mid  C= c).
\end{align*}
Likewise define the interval $\mathcal{I}^{g^*}_{b,c}$ to be the analogous interval constructed by minimizing and maximizing over $g^*\in\Pi^*$, and additionally conditioning on $B=b$. 

Suppose $Y^a \independent A \mid L$ for $a\in\{0,1\}$. Then, for each $c$, 
$$
    \mathcal{I}^{g^*}_{b,c} \equiv  \mathcal{I}^{g}_{c} \text{ for all } b.
$$
\end{proposition*}
The following proof relies on results from \citet{robins1986new}.
\begin{proof}
 If  $Y^a \independent A \mid L$, then for any $g^*\in\Pi^*$, $\mathbb{E}(Y^{g^*} \mid B=b, C=c)$ is equal to $\mathbb{E}(Y^{g^{\mathbf{rand}}} \mid B=b , C=c)$ where $g^{\mathbf{rand}}$ is some regime that is maximally dependent on $L$ and an exogenous randomizer term $\delta$. \citet{robins1986new} showed that $\mathbb{E}(Y^{g^{\mathbf{rand}}} \mid C=c)$ is equal to some convex combination of the values in the set $\Big\{\mathbb{E}(Y^g \mid C=c): g\in\Pi\Big\}$. The result then follows immediately.
\end{proof}
}

\subsubsection{Estimation proofs} \label{appsec: estimproofs}

%Efficient IF proposition
{
\renewcommand{\theproposition}{}
\begin{proposition*}[Re-statement of Proposition \ref{thm:proof eff infl}]
Suppose covariates $L$ are discrete. The efficient influence function of $\Psi(a,l)$  in $\mathcal{M}_{\text{np}}$ is given by 
\begin{align*}
    &  \Psi^{\text{eff}}(a,l) \color{black} = \\
    & \frac{1}{P( A = 1-a \mid  L = l)^2}\Bigg(\psi_1^{\text{eff}}(a,l)P( A = 1-a  \mid L = l)\\
    &- \psi_1(a,l) \frac{I[L = l]}{P(L = l)} (I[A = 1-a] - P(A = 1-a  \mid  L = l))\\
    & - \bigg(\frac{I[A = a, L = l]}{P(A = a ,L = l)}(Y - \mathbb{E}(Y  \mid  A = a, L = l)) P(A = a  \mid  L = l)  \\
    & + \mathbb{E}(Y  \mid  A = a, L = l)\frac{I[L = l]}{P(L = l)}(I[ A= a] - P( A = a  \mid  L = l))\bigg)P(A = 1-a  \mid  L = l) \\
    & + \mathbb{E}(Y  \mid  A = a, L = l) P( A = a  \mid  L = l) \frac{I[L = l]}{P(L = l)} (I[A = 1-a] - P(A = 1-a  \mid  L = l)) \Bigg),
\end{align*}
where 
\begin{align*}
 \psi_1^{\text{eff}}(a,l) =  &\frac{(2Z-1) (2A - 1) Y I[A = a]}{\delta(l)f(Z \mid l)P(L = l)} - \bigg(\frac{Z \mathbb{E}((2A - 1) Y I[A = a] \mid Z,L= l)}{\delta(l)f(Z \mid l) P(L = l)} \nonumber \\
    & - \sum_z \frac{z \mathbb{E}((2A-1)YI[A = a] \mid Z = z, L = l)}{\delta(l)P(L = l)} \nonumber \\
    & + \frac{(2Z-1)[(2A-1) - \mathbb{E}(2A - 1 \mid Z,L=l)]}{2 f(Z \mid l) \delta(l)P(L = l)}\sum_z\frac{z \mathbb{E}((2A-1)YI[A = a] \mid Z = z, L = l)}{\delta(l)} \bigg) \nonumber \\
    & - \mathbb{E}\left(\frac{(2Z- 1) (2A - 1) Y I[A = a]}{\delta(l)f(Z \mid l)P(L = l)} \mid L = l  \right) ,
\end{align*}
that is, $ \psi_1^{\text{eff}}(a,l)$ is the nonparametric influence function for $\psi_1(a,l)$.
\end{proposition*}
The following proof uses the efficient influence function of $\psi_1$, which is derived in \citet{cui2021semiparametric}.
\begin{proof}[of Proposition \ref{thm:proof eff infl}]
To express the efficient influence function of $\Psi(a,l)$, we find the efficient influence function of (i) $\psi_1(a,l)=\mathbb{E}(Y^a  \mid  L = l)$, (ii) $\psi_2(a,l) = \mathbb{E}(Y\mid A=a, L = l) $ and (iii) $\psi_3(a,l) = P(A = a  \mid  L = l)$. Let $a' = 1 - a$. Using differentiation rules, the efficient influence function is 
\begin{align}
    \frac{\psi_1^{\text{eff}}(a) \psi_3(a' ) - \psi_3^{\text{eff}}(a')\psi_1(a) - \{\psi_2^{\text{eff}}(a) \psi_3(a) + \psi_2(a) \psi_3^{\text{eff}}(a)\}\psi_3(a') + \psi_2(a) \psi_3(a) \psi_3^{\text{eff}}(a')}{\psi_3^2(a')}, \label{eq: if derivatives}
\end{align}
where  we have slightly abused notation by omitting $l$ from all the arguments to make the expression less cluttered; for example we wrote $\psi_3(a' )$ instead of $\psi_3(a' ,l)$.
First, we will use that
\begin{align*}
\frac{d\psi_1(a,l; \theta_t)}{dt}\Big\vert_{t=0}
\end{align*}
was derived  in the supplementary material of \cite{cui2021semiparametric}; indeed, the nonparametric influence function for $\psi_1(a,l)$ is 

\begin{align*}
 & \psi_1^{\text{eff}}(a,l) \\
 =  &\frac{(2Z-1) (2A - 1) Y I[A = a]I[L = l]}{\delta(l)f(Z \mid l)P(L = l)} - \bigg(\frac{Z \mathbb{E}((2A - 1) Y I[A = a] \mid Z,L= l)}{\delta(l)f(Z \mid l) P(L = l)} \nonumber \\
    & - \sum_z \frac{z \mathbb{E}((2A-1)YI[A = a] \mid Z = z, L = l)}{\delta(l)P(L = l)} \nonumber \\
    & + \frac{(2Z-1)[(2A-1) - \mathbb{E}(2A - 1 \mid Z,L=l)]I[L = l]}{2 f(Z \mid l) \delta(l)P(L = l)}\sum_z\frac{z \mathbb{E}((2A-1)YI[A = a] \mid Z = z, L = l)}{\delta(l)} \bigg) \nonumber \\
    & - \mathbb{E}\left[\frac{(2Z- 1) (2A - 1) Y I[A = a]}{\delta(l)f(Z \mid l)P(L = l)} \mid L = l  \right]. 
\end{align*}

Furthermore, 
\begin{align*}
    \frac{d\psi_2(a,l; \theta_t)}{dt}\Big\vert_{t=0} &= \mathbb{E}\{YS({Y\mid A=a, L = l})\mid A=a, L = l \}
\\ & = \mathbb{E}\left[\left\{Y-\mathbb{E}(Y\mid A=a, L = l)\right\}S({Y\mid A=a, L = l})\mid A=a, L = l \right]
\\& = \mathbb{E}\left[\frac{I(A=a, L = l)\left\{Y-\psi_2(a,l) \right\}}{P(A=a,L = l)}S(\mathcal{O})\right]
\end{align*}
and, similarly,
\begin{align*}
    \frac{d\psi_3(a,l; \theta_t)}{dt}\Big\vert_{t=0} &= \mathbb{E}\left[\frac{I(L = l)}{P(L = l)}\{I(A = a) - \psi_3(a,l) \} S(\mathcal{O}) \right].
\end{align*}
By plugging in these derivatives into \eqref{eq: if derivatives}, and after some steps of simple algebra, we obtain the formula given in the statement.
\end{proof}
}

%Asymp normality proposition
{
\renewcommand{\theproposition}{}
\begin{proposition*}[Re-statement of Proposition \ref{Prop: Asympt Normal}]
Under Assumptions \ref{ass: cons}-\ref{ass: pos}, if
\begin{enumerate}
\item $Y$ is bounded, 
\item $\delta(L)$ is bounded away from $0$ w.p.1,
\item $\hat{P}(A = a | L = l) \in (\epsilon_1, 1-\epsilon_1)$ w.p.1 for some $\epsilon_1 > 0$,
\item $\hat{\delta(L)} \in (\epsilon_2, 1-\epsilon_2)$ w.p.1 for some $\epsilon_2 > 0$,
\item $\hat{\mathbb{E}}(Y | A,L), \hat{P}(A = a | Z, L), \hat{\mathbb{E}}(Y(2A-1)I(A = a) | L,Z)$, and $\hat{P}(A = a | L )$ are $P$-Donsker with vanishing variance, then 
\end{enumerate}

\begin{align*}
\sqrt{n}(\hat{\mathbb{E}}(Y^g) - \mathbb{E}(Y^g)) \to^d N(0,\sigma^2),
\end{align*}
for some $\sigma^2 < \infty$ as $n \to \infty$.

\end{proposition*}

To prove Proposition \ref{Prop: Asympt Normal}, we first introduce two lemmas.
\begin{lemma}
Let $g:\{0,1\} \times \mathcal{L} \to \{0,1\}$ be a function of $A$ and $L$. Then, under Assumptions \ref{ass: cons}-\ref{ass: pos},
\begin{align*}
\mathbb{E}(Y^g) &= \mathbb{E}\bigg( I[g(A,L) \neq A]V(A,L) + I[g(A,L) = A]\mathbb{E}(Y |A, L)\bigg),
\end{align*} 
where
\begin{align}
V(a,l) &:= \frac{\psi_1(1-a,l) - \mathbb{E}(Y | A = 1-a,L =l)P(A = 1-a | L = l)}{P(A = a| L = l)}. \label{eq: V definition}
\end{align}
\label{lemma: Id formula of E(Y^g)}
\end{lemma}
The following proof uses an identification formula under Assumptions \ref{ass: cons}-\ref{ass: pos} for $\mathbb{E}(Y^a | L = l)$ from \citet{cui2021semiparametric}.
\begin{proof}
Assumption \ref{ass: cons} implies $Y^g = I[g(A,L) \neq A]Y^{1-A} + I[g(A,L) = A]Y$. Then, by the law of total expectation,
\begin{align}
\mathbb{E}(Y^g) &= \mathbb{E}(\mathbb{E}(Y^g|A,L))
\nonumber \\
&= \mathbb{E}(\mathbb{E}(I[g(A,L) \neq A]Y^{1-A} + I[g(A,L) = A]Y | A,L)) \nonumber\\
&= \mathbb{E}(I[g(A,L) \neq A]\mathbb{E}(Y^{1-A}|A,L) + I[g(A,L) = A]\mathbb{E}(Y|A,L))  \label{eq: Algo Pf step 1}.
\end{align}
However, Lemma \ref{lemma: functional} implies that
\begin{align}
\mathbb{E}(Y^{1-a} |A = a, L) &= \frac{\mathbb{E}(Y^{1-a}|L) - \mathbb{E}(Y|A = 1-a,L)P(A = 1-a| L)}{P(A = a | L)}, \nonumber \\
&=\frac{\psi_1(1-a,L) - \mathbb{E}(Y|A = 1-a,L)P(A = 1-a| L)}{P(A = a | L)} = V(a,l), \label{eq: Algo Pf lemma 1}
\end{align}
where the second equality follows from \citet{cui2021semiparametric}, and the third from Equation \eqref{eq: V definition}.

Then, plugging Equation \eqref{eq: Algo Pf lemma 1} into Equation \eqref{eq: Algo Pf step 1} we obtain
\begin{align*}
\mathbb{E}(Y^g) &= \mathbb{E}\bigg( I[g(A,L) \neq A]V(A,L) + I[g(A,L) = A]\mathbb{E}(Y |A, L)\bigg).
\end{align*}
\end{proof}

Let $\hat{\mathbb{E}}(Y^g)$ denote the estimator described in the algorithm above. Then, $\hat{\mathbb{E}}(Y^g)$ is an asymptotically normal estimator of $\mathbb{E}(Y^g)$ if $\sqrt{n}(\hat{\mathbb{E}}(Y^g) - \mathbb{E}(Y^g)) \to^P N(0,\sigma^2)$ as $n \to \infty$, for some $0 < \sigma^2 < \infty$.
\begin{lemma}
Under the conditions of Lemma \ref{lemma: Id formula of E(Y^g)}, the difference
\begin{align*}
\hat{\mathbb{E}}(Y^g) - \mathbb{E}(Y^g) &= \textcircled{\raisebox{-0.05em}{\scalebox{0.85}{A}}} + \textcircled{\raisebox{-0.05em}{\scalebox{0.85}{B}}} + \textcircled{\raisebox{-0.05em}{\scalebox{0.85}{C}}} +
\textcircled{\raisebox{-0.05em}{\scalebox{0.85}{D}}},
\end{align*}
where 
\begin{align*}
\textcircled{\raisebox{-0.05em}{\scalebox{0.85}{A}}} &= \ \mathbb{P}_n I[g(A,L) \neq A]\bigg(\hat{V}(A,L) - V(A,L)\bigg),\\
\textcircled{\raisebox{-0.05em}{\scalebox{0.85}{B}}} &= \mathbb{P}_n I[g(A,L) \neq A] V(A,L) - \mathbb{E}(I[g(A,L) \neq A)V(A,L)],\\
\textcircled{\raisebox{-0.05em}{\scalebox{0.85}{C}}} &= \mathbb{P}_n I[g(A,L) = A]\bigg(\hat{\mathbb{E}}(Y |A, L ) -  \mathbb{E}(Y |A , L ) \bigg),\\
\textcircled{\raisebox{-0.05em}{\scalebox{0.85}{D}}} &= \mathbb{P}_n I[g(A,L) = A]\mathbb{E}(Y |A , L ) - \mathbb{E}(I[g(A,L) = A]\mathbb{E}(Y |A, L)).
\end{align*}
\label{lemma: Diff Decomp}
\end{lemma}
\begin{proof}
Step 5 of the estimation procedure and Lemma \ref{lemma: Id formula of E(Y^g)} imply
\begin{align*}
\hat{\mathbb{E}}(Y^g) - \mathbb{E}(Y^g) &= \mathbb{P}_n\bigg\{I[g(A,L) \neq A]\hat{V}(A,L) + I[g(A,L) = A]\hat{\mathbb{E}}(Y |A , L )\bigg\}\\
& - \mathbb{E}\bigg\{ I[g(A,L) \neq A]V(A,L) + I[g(A,L) = A]\mathbb{E}(Y |A, L)\bigg\},
\end{align*}
where 
\begin{align*}
\hat{V}(a,l) &:= \frac{\hat{\psi_1}(1-a,l) - \hat{\mathbb{E}}(Y | A = 1-a,L =l)\hat{P}(A = 1-a | L = l)}{\hat{P}(A = a| L = l)}.
\end{align*}
Therefore, 
\begin{align*}
\hat{\mathbb{E}}(Y^g) - \mathbb{E}(Y^g) &= \underbrace{\bigg(\mathbb{P}_n I[g(A,L) \neq A]\hat{V}(A,L) - \mathbb{E}(I[g(A,L) \neq A)V(A,L)] \bigg)}_{(\ast)}\\
&+ \underbrace{\bigg(\mathbb{P}_n I[g(A,L) = A]\hat{\mathbb{E}}(Y |A, L) - \mathbb{E}(I[g(A,L) = A]\mathbb{E}(Y |A, L)) \bigg)}_{(\ast\ast)}.
\end{align*}
However, if we add and subtract $\mathbb{P}_n I[g(A,L) \neq A]V(A,L)$ to $(\ast)$, we obtain
\begin{align*}
(\ast) &= \underbrace{\mathbb{P}_n (I[g(A,L) \neq A]\hat{V}(A,L) - I[g(A,L) \neq A]V(A,L))}_{\textcircled{\raisebox{-0.05em}{\scalebox{0.85}{A}}}} \\
&+ \underbrace{\mathbb{P}_n I[g(A,L) \neq A] V(A,L) - \mathbb{E}(I[g(A,L) \neq A)V(A,L)]}_{\textcircled{\raisebox{-0.05em}{\scalebox{0.85}{B}}}}.
\end{align*}
Furthermore, if we add and subtract $\mathbb{P}_n I[g(A,L) = A]\mathbb{E}(Y |A, L)$ to $(\ast\ast)$, we obtain
\begin{align*}
& (\ast\ast) \\
&= \underbrace{\mathbb{P}_n I[g(A,L) = A]\hat{\mathbb{E}}(Y |A, L) -  I[g(A,L) = A]\mathbb{E}(Y |A, L)}_{\textcircled{\raisebox{-0.05em}{\scalebox{0.85}{C}}}}\\
& + \underbrace{\mathbb{P}_n I[g(A,L) = A]\mathbb{E}(Y |A, L) - \mathbb{E}(I[g(A,L) = A]\mathbb{E}(Y |A, L))}_{\textcircled{\raisebox{-0.05em}{\scalebox{0.85}{D}}}},
\end{align*}
which concludes the proof.
\end{proof}

Term \textcircled{\raisebox{-0.05em}{\scalebox{0.85}{A}}} in Lemma \ref{lemma: Diff Decomp} verifies
\begin{align*}
&\textcircled{\raisebox{-0.05em}{\scalebox{0.85}{A}}}\\
&= \mathbb{P}_n I[g(A,L) \neq A]\bigg(\hat{V}(A,L) - V(A,L)\bigg),\\
&= \mathbb{P}_n I[g(A,L) \neq A]\bigg(\frac{\hat{\psi_1}(1-A,L) - \hat{\mathbb{E}}(Y |  1-A,L)\hat{P}(A = 1-A | L)}{\hat{P}(A| L)}\\
&- \frac{\psi_1(1-A,L) - \mathbb{E}(Y |  1-A,L)P(A = 1-A | L)}{P(A| L)}\bigg),\\
&= \mathbb{P}_n \frac{I[g(A,L) \neq A]}{P(A| L) \hat{P}(A| L)}\\
&\cdot \bigg(\underbrace{(\hat{\psi_1}(1-A,L)P(A| L) - \psi_1(1-A,L)\hat{P}(A| L))}_{(\dagger)}\\
&\underbrace{\begin{aligned}
& + (\mathbb{E}(Y | 1-A,L)P( 1-A | L)\hat{P}(A| L)\\
& - \hat{\mathbb{E}}(Y |  1-A,L)\hat{P}(1-A | L)P(A| L))\bigg) \end{aligned}}_{(\ddag)}.\\
\end{align*}
However if we add and subtract $\psi_1(1-A,L)P(A| L)$ to $(\dagger)$,
\begin{align}
(\dagger) &= (\hat{\psi_1}(1-A,L) - \psi_1(1-A,L))P(A| L) 
 +\psi_1(1-A,L)(P(A| L) - \hat{P}(A| L)). \label{eq: dagger eq}
\end{align}

Similarly, if we add and subtract $\mathbb{E}(Y | 1-A,L)P(A = 1-A | L)P(A| L)$ to $(\ddag)$, 
\begin{align}
&(\ddag) = \mathbb{E}(Y | 1-A, L)P(1-A | L)(\hat{P}(A| L) - P(A| L)) + P(A| L) \cdot (\dagger \dagger), \label{eq: ddag eq}
\end{align}
where 
\begin{align*}
    (\dagger \dagger) &= \mathbb{E}(Y | 1-A,L)P(A = 1-A | L) - \hat{\mathbb{E}}(Y |  1-A,L)\hat{P}(A = 1-A | L).
\end{align*}
Furthermore, adding and subtracting $\mathbb{E}(Y | 1-A,L)\hat{P}( 1-A | L)$ to $(\dagger\dagger)$,
\begin{align}
(\dagger\dagger) &= \mathbb{E}(Y |  1-A,L)(P(1-A | L) -\hat{P}(1-A | L)) \nonumber\\
&+ (\mathbb{E}(Y |  1-A,L)  - \hat{\mathbb{E}}(Y |  1-A,L))\hat{P}( 1-A | L). \label{eq: dagger dagger eq}
\end{align}
We can now prove Proposition \ref{Prop: Asympt Normal}.

\begin{proof}
By Lemma \ref{lemma: Diff Decomp},
\begin{align*}
\sqrt{n}(\hat{\mathbb{E}}(Y^g) - \mathbb{E}(Y^g)) = \sqrt{n}(\textcircled{\raisebox{-0.05em}{\scalebox{0.85}{A}}} + \textcircled{\raisebox{-0.05em}{\scalebox{0.85}{B}}} + \textcircled{\raisebox{-0.05em}{\scalebox{0.85}{C}}} +
\textcircled{\raisebox{-0.05em}{\scalebox{0.85}{D}}}).
\end{align*}
We look at each term of the difference (multiplied by $\sqrt{n}$) individually. 
\begin{align*}
\sqrt{n} \times \textcircled{\raisebox{-0.05em}{\scalebox{0.85}{A}}} &= \sqrt{n}\mathbb{P}_n \frac{I[g(A,L) \neq A]}{P(A| L) \hat{P}(A| L)} [(\dagger) + (\ddag)],\\
&= \mathbb{P}_n \frac{I[g(A,L) \neq A]}{P(A| L) \hat{P}(A| L)} [\sqrt{n}\times(\dagger) + \sqrt{n}\times(\ddag)].
\end{align*}
However, multiplying Equation \eqref{eq: dagger eq} by $\sqrt{n}$,  we obtain
\begin{align*}
\sqrt{n} \times (\dagger) &= P(A | L) \times \sqrt{n}(\hat{\psi}_1(1-A, L) - \psi_1(1-A,L))\\
& + \psi_1(1-A,L) \times \sqrt{n}(P(A | L) - \hat{P}(A | L)),
\end{align*}
and
\begin{align*}
&\sqrt{n} (\hat{\psi}_1(a, l) - \psi_1(a,l))\\
&= \sqrt{n} \bigg(\sum_z \frac{(2z-1)}{\hat{\delta}(l)}\hat{\mathbb{E}}(Y(2A-1)I(A = a)| L = l, Z = z)\\
&- \mathbb{E}\bigg[\frac{(2Z-1)Y(2A-1)I(A=a)}{\delta(L)f(Z|L)} | L = l\bigg] \bigg)\\
&= \sqrt{n} \sum_z \bigg(\frac{(2z-1)}{\hat{\delta}(l)}\hat{\mathbb{E}}(Y(2A-1)I(A = a)| L = l, Z = z)\\
&- \frac{(2z-1)}{\delta(l)}\mathbb{E}(Y(2A-1)I(A = a)| L = l, Z = z) \bigg)\\
&=  \sum_z \frac{(2z-1)}{\hat{\delta}(l)\delta(l)}\bigg[\delta(l)\sqrt{n}\bigg(\hat{\mathbb{E}}(Y(2A-1)I(A = a)| L = l, Z = z)\\
&- \mathbb{E}(Y(2A-1)I(A = a)| L = l, Z = z)\bigg)\\
& + \mathbb{E}(Y(2A-1)I(A = a)| L = l, Z = z)\sqrt{n}(\delta(l) - \hat{\delta}(l)\bigg]\\
&=  \sum_z \frac{(2z-1)}{\hat{\delta}(l)\delta(l)}\bigg[\delta(l)\sqrt{n}\bigg(\hat{\mathbb{E}}(Y(2A-1)I(A = a)| L = l, Z = z)\\
&- \mathbb{E}(Y(2A-1)I(A = a)| L = l, Z = z)\bigg)\\
& + \mathbb{E}(Y(2A-1)I(A = a)| L = l, Z = z)\bigg(\sqrt{n}(P(A= 1 |Z = 1,L)  - \hat{P}(A = 1 | Z = 1, L))\\
& + \sqrt{n}(P(A= 1 |Z = 0,L)  - \hat{P}(A = 1 | Z = 0, L)  \bigg) \bigg].
\end{align*}

Furthermore, plugging Equation \eqref{eq: dagger dagger eq} in Equation \eqref{eq: ddag eq} and multiplying by $\sqrt{n}$, we obtain
\begin{align*}
&\sqrt{n} \times (\ddag)\\
&= \mathbb{E}(Y |  1-A, L)P( 1-A | L)\\
&\times \sqrt{n}(\hat{P}(A| L) - P(A| L))\\
&+P(A | L)\mathbb{E}(Y | 1-A,L)\\
&\times \sqrt{n}(P(A = 1-A | L) -\hat{P}(A = 1-A | L))\\
&+P(A | L)\hat{P}(A = 1-A | L)\\
&\times \sqrt{n}(\mathbb{E}(Y |  1-A,L)  - \hat{\mathbb{E}}(Y | 1-A, L)).
\end{align*}
Then, Assumptions \ref{Ass: Bounded Y}-\ref{Ass: Donsker prop} of Proposition \ref{Prop: Asympt Normal} imply that $\sqrt{n} \times \textcircled{\raisebox{-0.05em}{\scalebox{0.85}{A}}} \to 0$ as $n \to \infty$.

Furthermore, by the central limit theorem, 
\begin{align*}
&\sqrt{n} \times (\textcircled{\raisebox{-0.05em}{\scalebox{0.85}{B}}} + \textcircled{\raisebox{-0.05em}{\scalebox{0.85}{D}}})\\
&= \sqrt{n}(\mathbb{P}_n (I[g(A,L) \neq A] V(A,L) + I[g(A,L) = A]\mathbb{E}(Y |A, L))\\
& - \mathbb{E}(I[g(A,L) \neq A)V(A,L) + I[g(A,L) = A]\mathbb{E}(Y |A, L)])\\
& \to^d N(0, \sigma^2),
\end{align*}
for some $\sigma^2 < \infty$.

Finally,
\begin{align*}
\sqrt{n} \times \textcircled{\raisebox{-0.05em}{\scalebox{0.85}{C}}} &= \mathbb{P}_n I[g(A,L) = A]\sqrt{n}\bigg(\hat{\mathbb{E}}(Y |A, L) -  \mathbb{E}(Y |A, L) \bigg),
\end{align*}
converges to $0$ by Assumption \ref{Ass: Donsker prop} of Proposition \ref{Prop: Asympt Normal}.

Hence,
\begin{align*}
\sqrt{n}(\hat{\mathbb{E}}(Y^g) - \mathbb{E}(Y^g)) \to^d N(0,\sigma^2),
\end{align*}
for some $\sigma^2 < \infty$.
\end{proof}
}

\subsubsection{Proof of Proposition \ref{prop: reform}} \label{appsec: Prop3proof}
Consider the set of joint distributions of $(Y^{a=1}, Y^{a=0}, A)$ conditional on $L=l$. We could partition this set according to the following 4 mutually-exclusive conditions, where $\tau_l(a') \coloneqq \mathbb{E}(Y^{a=1}  \mid A=a', L=l) - \mathbb{E}(Y^{a=0}  \mid A=a', L=l)$:

    \begin{itemize}
        \item [(\textbf{1.})] $\tau_l(1)\geq0$ and $\tau_l(0)\geq0$,
        \item [(\textbf{2.})] $\tau_l(1)<0$ and $\tau_l(0)<0$,
        \item [(\textbf{3.})] $\tau_l(1)\geq0$ and $\tau_l(0)<0$,
        \item [(\textbf{4.})] $\tau_l(1)<0$ and $\tau_l(0)\geq0$.
    \end{itemize}

   If condition (\textbf{3.}) holds, then $g_{\mathbf{sup}}(A, l) = A$. If condition (\textbf{4.}) holds, then $g_{\mathbf{sup}}(A, l) = 1-A$. 

Let $\tau_l^*\coloneqq \mathbb{E}(Y^{a=1}  \mid L=l) - \mathbb{E}(Y^{a=0}  \mid  L=l)$. The following identity  follows by laws of probability,

\begin{align*}
    \mathbb{E}(Y^a \mid L=l) = \mathbb{E}(Y^{a}  \mid A=1, L=l)P(A=1 \mid L=l) + \mathbb{E}(Y^{a}  \mid A=0, L=l)P(A=0 \mid L=l).
\end{align*}

Thus, if condition (\textbf{1.}) holds, then $g_{\mathbf{sup}}(A, l) = 1$ and also $g_{\mathbf{opt}}(l) = 1$ and so $g_{\mathbf{sup}}(A, l) = g_{\mathbf{opt}}(l)$. Likewise, if condition (\textbf{2.}) holds, then $g_{\mathbf{sup}}(A, l) = 0$ and also $g_{\mathbf{opt}}(l) = 0$ and so $g_{\mathbf{sup}}(A, l) = g_{\mathbf{opt}}(l)$. 

Let $\gamma$ be the law-dependent function defined as

\begin{align*}
\gamma(l) = 
    \begin{cases}
        0 & \text{ if } \{\tau_l(1)\geq0, \tau_l(0)\geq0\} \text{ or } \{\tau_l(1)<0, \tau_l(0)<0\} \\ 
        1 & \text{ if } \{\tau_l(1)\geq0, \tau_l(0)<0\} \\
        2 & \text{ if } \{\tau_l(1)<0, \tau_l(0)\geq0\}.
    \end{cases}
\end{align*}

Then, Proposition \ref{prop: reform} follows immediately. Furthermore, because $g_{\mathbf{sup}}$ and $\gamma$ are each identified whenever $\tau_l(a)$ is identified for all $a,l$, it follows that $\gamma$ is identified whenever $g_{\mathbf{sup}}$ is identified.

\end{document}